\documentclass[aps,preprint,preprintnumbers,superscriptaddress,amsmath,amssymb,prx]{revtex4-2}

\usepackage{graphicx}
\usepackage{color}
\usepackage{comment}
\usepackage[normalem]{ulem}
\usepackage{bm}
\usepackage{cases}

\renewcommand{\bm}[1]{{\mbox{\boldmath $#1$}}}

\begin{document}

\title{Chiral superconductivity in UTe$\bm{_2}$ probed by anisotropic low-energy excitations}

\author{K.~Ishihara}\email{ishihara@qpm.k.u-tokyo.ac.jp}
\author{M.~Roppongi}
\author{M.~Kobayashi}
\author{Y.~Mizukami}
\affiliation{Department of Advanced Materials Science, University of Tokyo, Kashiwa, Chiba 277-8561, Japan}

\author{H.~Sakai}
\author{Y.~Haga}
\affiliation{Advanced Science Research Center, Japan Atomic Energy Agency, Tokai, Ibaraki 319-1195, Japan}

\author{K.~Hashimoto}
\author{T.~Shibauchi}\email{shibauchi@k.u-tokyo.ac.jp}
\affiliation{Department of Advanced Materials Science, University of Tokyo, Kashiwa, Chiba 277-8561, Japan}

\date{\today}
\maketitle


{\bf
Chiral spin-triplet superconductivity is a topologically nontrivial pairing state with broken time-reversal symmetry, which can host Majorana quasiparticles~\cite{Kallin2016,Kozii2016}. The recently discovered heavy-fermion superconductor UTe$\bm{_2}$~\cite{Ran2019_S} exhibits peculiar properties of spin-triplet pairing~\cite{Ran2019_S,Aoki2019,Ran2019_NP,Nakamine2021}, and the possible chiral state has been actively discussed~\cite{Jiao2020,Hayes_arxiv,Bae2021}. However, the symmetry and nodal structure of its order parameter in the bulk, which determine the Majorana surface states~\cite{Kozii2016}, remains controversial~\cite{Metz2019,Ishizuka2021,Nevidomskyy_arxiv,Shishidou2021}. Here we focus on the number and positions of superconducting gap nodes in the ground state of UTe$\bm{_2}$. Our magnetic penetration depth measurements for three field orientations in the Meissner state reveal the power-law temperature dependence with exponents nearly equal to 2 or less than 2, which excludes single-component spin-triplet states. The anisotropy of low-energy quasiparticle excitations indicates multiple point nodes near the $\bm{k_y}$- and $\bm{k_z}$-axes, evidencing that the order parameter has multiple components in a chiral complex form. We find that most consistent is a chiral $\bm{B_{3u}+iA_u}$ non-unitary state, which provides fundamentals of the topological properties in UTe$\bm{_2}$.
}
\clearpage


Since the discovery of superconductivity in the nonmagnetic uranium-based compound UTe$_2$, which locates near a ferromagnetic quantum critical point\,\cite{Ran2019_S}, the nature of its superconducting state has been extensively studied\,\cite{Aoki2019,Ran2019_NP,
Nakamine2021,Knebel2019,
Jiao2020,Hayes_arxiv,Bae2021,Metz2019,Kittaka2020,Thomas2020,Braithwaite2019,
Aoki2020,Haga_arxiv,
Thomas2021,Rosa_arxiv,Miao2020,
Ishizuka2019,Nevidomskyy_arxiv,
Ishizuka2021,Shishidou2021,Paulsen_arxiv}. Recent studies have reported several anomalous superconducting properties, including extremely high upper critical field much beyond the Pauli limit\,\cite{Ran2019_S,Aoki2019,Ran2019_NP}, reentrant superconductivity\,\cite{Ran2019_NP,Knebel2019
}, and little reduction of the Knight shift in the nuclear magnetic resonance (NMR)\,\cite{Ran2019_S,
Nakamine2021}. These results suggest odd-parity spin-triplet pairing in UTe$_2$, as in the case of uranium-based ferromagnetic superconductors\,\cite{Aoki_review}. The symmetry of the superconducting order parameter is closely related to the superconducting gap structure, and the previous studies of low-energy quasiparticle excitations, such as specific heat, thermal transport, and magnetic penetration depth measurements\,\cite{Bae2021,Metz2019,Kittaka2020}, support the presence of point nodes, consistent with the spin-triplet pairing states. 

More intriguingly, recent scanning tunneling microscopy\,\cite{Jiao2020}, optical Kerr effect\,\cite{Hayes_arxiv}, and microwave surface impedance measurements\,\cite{Bae2021} suggest time-reversal symmetry breaking (TRSB) in the superconducting state at ambient pressure. As the sign of imaginary part changes under the time-reversal transformation, the chiral TRSB state requires multiple order parameter components in complex form. We note that high-pressure studies reveal several superconducting phases\,\cite{Thomas2020,
Braithwaite2019,Aoki2020
}, suggesting the presence of multiple order parameters under pressure. Thus, UTe$_2$ is a prime candidate of a topological chiral spin-triplet superconductor. However, the symmetry of the odd-parity vector order parameter $\bm{d}$, whose magnitude is the gap size and whose direction is perpendicular to the spins of Cooper pairs, is still highly controversial. 
Especially, the nodal structure of order parameter and whether or not it is chiral in the ground state are important issues to understand the possible topological properties of UTe$_2$.

\begin{figure*}[tbp]
    \includegraphics[width=\linewidth]{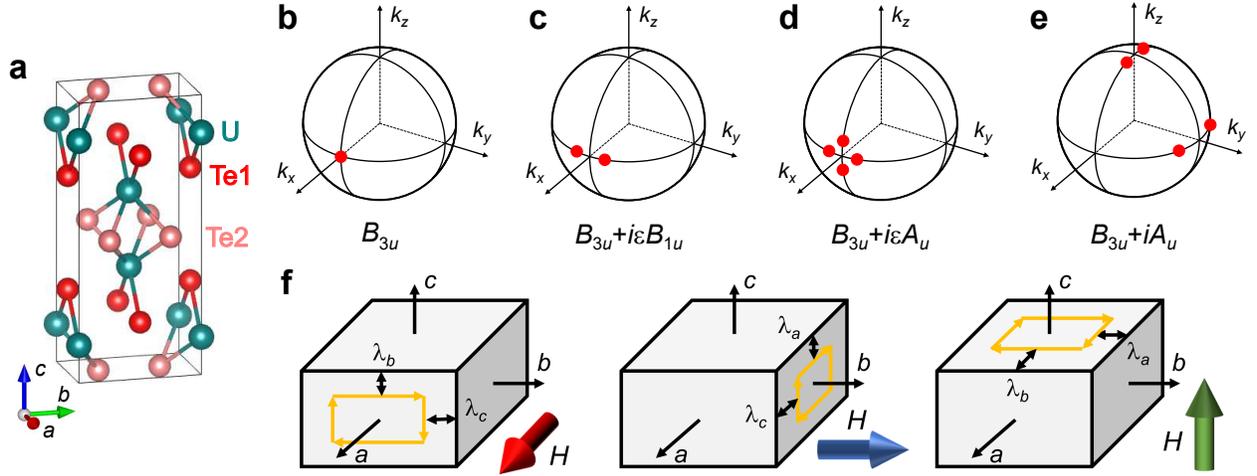}
    \caption{
       {\bf Nodal positions of several pairing states in UTe$_2$ and geometries of anisotropic penetration depth measurements.} {\bf a}, Crystal structure of UTe$_2$. {\bf b-e}, Positions of the point nodes (red points) for $B_{3u}$ ({\bf b}), $B_{3u}+i\varepsilon B_{1u}$ ({\bf c}), $B_{3u}+i\varepsilon A_u$ ({\bf d}), and $B_{3u}+iA_u$ ({\bf e}) order parameters, where $\varepsilon$ is a sufficiently small real number. {\bf f}, Schematic relations between the directions of the ac magnetic field (big arrows) and $\lambda$ components.}
    \label{F1}
\end{figure*}

The crystal structure of UTe$_2$ (Fig.\,\ref{F1}a) is classified into the point group $D_{2h}$, whose irreducible representations (IRs) of odd-parity order parameters are listed in Table\,\ref{T1}. 
In the cases of $B_{1u}$, $B_{2u}$, and $B_{3u}$ states, point nodes in the superconducting gap function exist on the $k_z$-, $k_y$-, and $k_x$-axes (Fig.\,\ref{F1}b), respectively, while the $A_u$ state is fully gapped. When the odd-parity order parameter ${\bm d}$ is represented by a single IR, the positions of nodes can be detected by the temperature dependence of the change in magnetic penetration depth, $\Delta \lambda (T)\equiv \lambda (T)-\lambda (0)$. This is because the low-temperature superfluid density $\lambda^{-2}(T)$, which is determined by thermally-excited quasiparticles near the nodes, depends strongly on the directions of the shielding supercurrent density $\bm{j}_{\rm s}$ and the point nodes (whose direction is defined as $\bm{I}$). As a result, when the point nodes are directed along crystallographic $\alpha$ axis ($\bm{I}\parallel \alpha$), $\Delta \lambda_\alpha (T)$ follows $T^2$ dependence, while for perpendicular axes $\beta$ and $\gamma$, $\Delta \lambda_\beta (T)$ and $\Delta \lambda_\gamma (T)$ should follow $T^4$ dependence (Table\,\ref{T1})\,\cite{Gross1986}. Here, the subscript $i$ of $\lambda_i$ represents the direction of supercurrent density $\bm{j}_{\rm s}$. 

On the other hand, when two symmetries in different IRs accidentally admix to form a TRSB complex order parameter, point nodes are generally located away from the high symmetry axes, and various nodal structures become possible\,\cite{Hayes_arxiv,Shishidou2021}. Considering the $B_{3u}+i\varepsilon B_{1u}$ ($B_{3u}+i\varepsilon A_u$) state, for example, where $\varepsilon$ is a sufficiently small real number, a point node of the $B_{3u}$ state splits into two (four) point nodes as depicted in Fig.\,\ref{F1}c (Fig.\,\ref{F1}d) (for more details, see Supplementary Information I). These split point nodes can be identified as topological Weyl nodes defined by a Chern number\,\cite{Hayes_arxiv,Shishidou2021}, and corresponding Majorana arc surface states are expected\,\cite{
Kozii2016}. Although an experimental determination of the exact positions of the nodes is quite challenging in these cases, we can summarize expected nodal positions for different relative sizes of two components in the complex order parameters in Table\,\ref{T2}. Thus, by detecting the anisotropy of quasiparticle excitations through direction-dependent physical quantities, such as $\Delta \lambda_i (T)$, we can pin down the superconducting symmetry among the non-chiral and chiral states listed in Tables\,\ref{T1} and \ref{T2}, respectively.


We use three independent measurements of resonant frequency of the tunnel-diode oscillator (see Methods) with weak ac magnetic field along the $a$-, $b$-, and $c$-axes, in which the shielding current flows perpendicular to the field as described in Fig.\,\ref{F1}f. Thus the frequency shift $\Delta f (T)$ consists of two penetration depth components perpendicular to the field direction. As a result, in the single component order parameter cases for point nodes along the $\alpha$ direction, $\Delta f (T)$ for $H_\omega \parallel \alpha$ is the sum of $\Delta \lambda_\beta (T)$ and $\Delta \lambda_\gamma (T)$ components and thus follows $T^4$ dependence, while $\Delta f (T)$ for $H_\omega \perp \alpha$ should follow $T^2$ dependence at low $T$. We stress that in our measurements the sample is in the Meissner state. Therefore, our approach is an ideal way to investigate the superconducting symmetry in the ground state in the zero-field limit at ambient pressure.


Figures\,\ref{F2}a, b show $\Delta f(T)$ in two single crystals of UTe$_2$ denoted as \#A1 and \#B1, respectively (see Supplementary Information V). We observe a large change in $\Delta f(T)$ at 1.65\,K (\#A1) and 1.75\,K (\#B1) corresponding to the superconducting transition. The clear superconducting transition at $T_c=1.65$\,K (1.75\,K) is also reproduced in the specific heat data of crystal \#A1 (\#B1) (Fig.\,\ref{F2}c). For crystal \#A1, in addition to the large change at 1.65K, we notice a small diamagnetic signal in $\Delta f(T)$ data and a small increase in the specific heat below $T_c^+\sim 1.9$\,K, while the data of crystal \#B1 seem to consist of only a single jump (Fig.\,\ref{F2}d). 
There are two possible origins of the anomaly at 1.9\,K for crystal \#A1. The first possibility is that a small distribution of chemical stoichiometry\,\cite{Haga_arxiv,Thomas2021} and/or local strains may induce a small region with higher $T_c$. The second is caused by the accidental coexistence of two order parameter components in the different IRs having slightly different critical temperatures $T_c$ and $T_c^+$, forming a TRSB chiral superconducting state below $T_c$.
We note that our specific heat data showing a high-$T$ kink and a low-$T$ jump are similar to the ones under pressure\,\cite{Thomas2020,
Braithwaite2019} and consistent with some of the ambient-pressure results, which may suggest the intrinsic nature of the double transitions\,\cite{Hayes_arxiv
}. Here, we emphasize that the observations of the two specific heat anomalies in crystal \#A1 and a single jump in crystal \#B1 do not necessarily contradict nearly the same weight of two order parameter components discussed later, because in the Landau theory the jump heights have nontrivial  dependence on the coefficients of the fourth power terms of the free energy for chiral superconducting order parameters \cite{Thomas2020,Sigrist1991}. However, a recent study\,\cite{Rosa_arxiv} has reported that the presence or absence of the double transitions depends strongly on the crystal growth conditions, and the origin of the double transitions is still highly controversial. Thus, our specific heat data cannot be taken as decisive evidence for a TRSB state, and we therefore rather focus on the low-temperature penetration depth data to discuss the gap symmetry in the superconducting state.

\begin{figure}[tbp]
    \includegraphics[width=\linewidth]{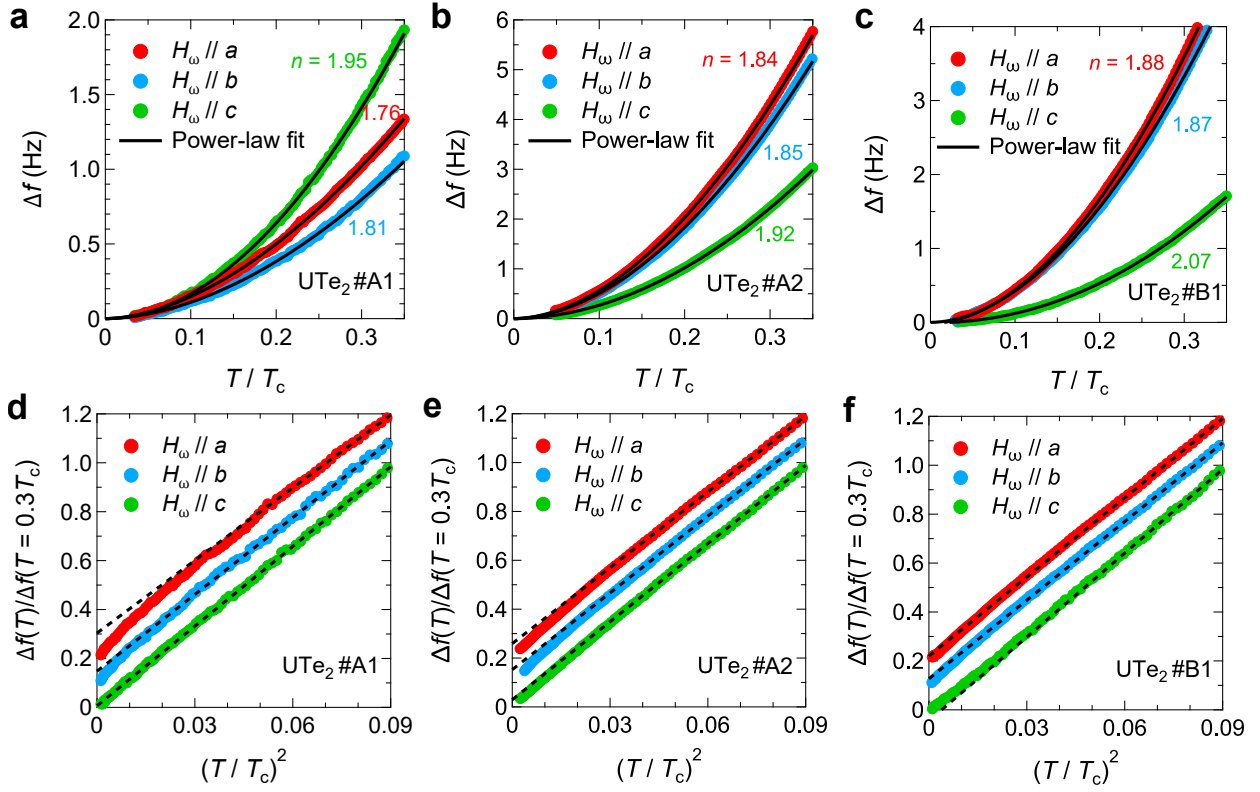}
    \caption{{\bf Anisotropic frequency shift in UTe$_2$ for three field directions.} {\bf a-c},  Low-$T$ behavior of $\Delta f$ in UTe$_2$ for crystal \#A1 ({\bf a}), \#A2 ({\bf b}), and \#B1 ({\bf c}) as a function of $T$ normalized by $T_c$. Solid lines represent the fitting curves with the power-law function. {\bf d-f}, $\Delta f (T)$ normalized by the value at $T=0.3T_c$ as a function of $(T/T_c)^2$ in UTe$_2$ for crystals \#A1 ({\bf {d}}), \#A2 ({\bf {e}}), and \#B1 ({\bf f}). Dashed lines represent $T^2$ dependence. The data with the field along the $a$- and $b$-axes are vertically shifted by 0.2 and 0.1, respectively.}
    \label{F3}
\end{figure}

The key results are the temperature dependence of $\Delta f(T)$ at low $T$, which is shown for crystals \#A1, \#A2, and \#B1 in Figs.\,\ref{F3}a-c, respectively. The black solid lines represent fitting curves for the data below $0.3T_c$ using the power-law function, $\Delta f (T) \propto T^n$. We note that the exponent values $n=1.95$ (\#A1), 1.95 (\#A2), and 2.07 (\#B1) for $H_\omega \parallel c$ are consistent with the previous $ab$-plane penetration depth studies\,\cite{Metz2019,Bae2021}. From the fittings, we find that the obtained exponent values for all field directions are nearly equal to 2 or less than 2 in all the samples. This feature can be more clearly seen by plotting $\Delta f(T)$ as a function of $(T/T_c)^2$ as shown in Figs.\,\ref{F3}d-f, where all data show almost linear or convex downward curvatures at low $T$. Thus, our results contradict any cases of single-component odd-parity order parameters, in which $\Delta f(T)$ should follow $T^4$ dependence when the applied field is directed to the point node direction. Another feature of our data is that the exponent values obtained from $\Delta f (T)$ for $H_\omega\parallel a$ and $H_\omega\parallel b$ are smaller than that for $H_\omega\parallel c$ in all samples. Considering that $\Delta f(T)$ consists of two $\Delta \lambda_i (T)$ components perpendicular to the magnetic field (Fig.\,\ref{F1}f), our exponent analysis indicates that the exponent value of $\Delta \lambda_c (T)$ is smaller than those of $\Delta \lambda_a (T)$ and $\Delta \lambda_b (T)$, which will be discussed in more detail below.

For further investigations of the gap structure, we extract $\Delta \lambda_i (T)$ separately from the $\Delta f(T)$ data for three different field orientations for crystals \#A1 and \#B1, by considering the geometry of the sample (see Supplementary Information VI). Such an analysis is valid when the magnetic penetration depth is much shorter than the sample dimensions, which holds at low temperatures. To compare the quasiparticle excitations along each crystallographic axis, we discuss the normalized superfluid density $\rho_{{\rm s},i}(T)=\lambda_i^2(0)/\lambda_i^2(T)$ and normalized penetration depth $\Delta \lambda_i (T)/\lambda_i (0)$ for three supercurrent directions $i =a$, $b$, and $c$, in which evaluations of $\lambda_i (0)$ values are needed. The anisotropy of $\lambda (0)$ can be estimated by the anisotropy of coherence length $\xi$ which can be determined from the initial slope of the temperature dependence of upper critical field $H_{\rm c2}(T)$, when we simply ignore the anisotropy of gap function (see Supplementary Information X). From the $H_{\rm c2}(T)$ data in Ref\,\cite{Ran2019_S}, we estimate $\lambda_a (0):\lambda_b (0): \lambda_c (0)=\xi_a^{-1}(T_c):\xi_b^{-1}(T_c):\xi_c^{-1}(T_c)=3.33:1:2.66$. By using the value $\sqrt{\lambda_a (0) \lambda_b (0)}\approx1$\,$\mu$m estimated from the previous penetration depth studies\,\cite{Metz2019,Bae2021}, we obtain $\lambda_a (0)=1820$\,nm, $\lambda_b (0)=550$\,nm, and $\lambda_c (0)=1460$\,nm.

\begin{figure}[t]
    \includegraphics[width=0.7\linewidth]{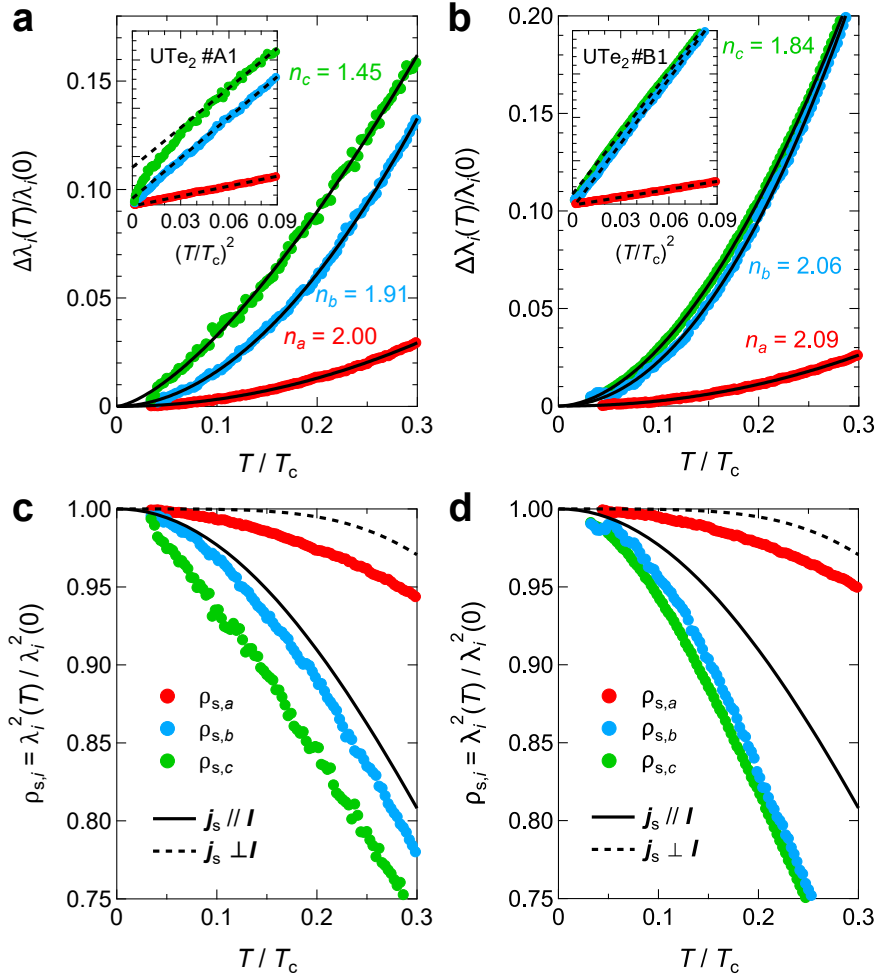}
    \caption{{\bf Anisotropic penetration depth in UTe$_2$ for three supercurrent directions.} {\bf a,b}, $\Delta \lambda_i (T)$ normalized by $\lambda_i (0)$ for $i=a$, $b$, and $c$ as a function of $T/T_c$ for crystals \#A1 ({\bf a}) and \#B1 ({\bf b}). Solid lines in {\bf a} ({\bf b}) represent the fitting curves of the power-law function with the exponents $n_a=2.00$ (2.09), $n_b=1.91$ (2.06), and $n_c=1.45$ (1.84). The inset shows the same data plotted against $(T/T_c)^2$, and the black dashed lines represent $T^2$ dependence. {\bf c,d}, Normalized superfluid density calculated from $\Delta \lambda_i (T)/\lambda_i (0)$ data for crystals \#A1 ({\bf c}) and \#B1 ({\bf d}). Solid and dashed lines represent the theoretical curves for the single order parameter case for ${\bm j}_{\rm s}$ parallel and perpendicular to the nodal direction ${\bm I}$, respectively.}
    \label{F4}
\end{figure}

The obtained normalized $\Delta \lambda_i (T)/\lambda_i (0)$ as a function of $T/T_c$ for three directions of crystals \#A1 and \#B1 are shown in Figs.\,\ref{F4}a,b, respectively. First of all, the exponent values $n_i$ obtained from the power-law fitting in $\Delta \lambda_i (T)/\lambda_i (0)$ data are all nearly equal to 2 or less than 2, which is again inconsistent with all the cases of the single component order parameter. Especially, as expected from $\Delta f(T)$ data, $n_c=1.45$ in crystal \#A1 and $n_c=1.84$ in crystal \#B1 are smaller than $n_a \approx n_b \approx 2$. The relatively small $n_c$ in crystal \#A1 can be more clearly seen by plotting the data as a function of $(T/T_c)^2$ (Fig.\,\ref{F4}a, inset). We note that the above discussions are independent on the fitting range of the power-law dependence of $\Delta \lambda (T)\propto T^n$ (see Supplementary Information VII). Another consequence of $\Delta \lambda (T)/\lambda (0)$ results is that the quasiparticle excitations along the $b$- and $c$-axes are much larger than those along the $a$-axis, implying a highly anisotropic nodal structure.
Figures\,\ref{F4}c,d show the normalized superfluid density, $\rho_{{\rm s,}i} \equiv \lambda_i^2 (T)/\lambda_i^2 (0)$, along each crystallographic axis for crystals \#A1 and \#B1, respectively, plotted against $T/T_c$. Compared with theoretical curves for the single order parameter with the supercurrent density $\bm{j}_s$ parallel and perpendicular to the direction of point nodes $\bm{I}$, the amount of the excitations along the $a$-axis is between the ${\bm j}_{\rm s} \parallel {\bm I}$ and ${\bm j}_{\rm s} \perp {\bm I}$ cases, while those along the $b$- and $c$-axes are even larger than the ${\bm j}_{\rm s} \parallel {\bm I}$ case.


Having established that our anisotropic superfluid density data exclude the single-component odd-parity order parameters, we now discuss the superconducting gap structure based on our experimental results. The multi-component order parameter can appear below $T_c$ if the two different IR components have close transition temperatures $T_c$ and $T_c^+(\ge\, T_c)$ accidentally. Adding two order parameters with preserving time-reversal symmetry will not split the point nodes and cannot account for our results (see Supplementary Information II), and thus we need to consider chiral superconducting states formed by two order parameters in different IRs (Table\,\ref{T2}). The closeness of the two transition temperatures found in our results as well as in several previous papers including ambient pressure\,\cite{Hayes_arxiv,Thomas2020
} and high pressure\,\cite{Thomas2020,
Braithwaite2019,Aoki2020
} studies suggests that the magnitudes of the two components are also close to each other. Based on these considerations and the observed large quasiparticle excitations along the $b$- and $c$-axes, we conclude that the $B_{3u}+iA_u$ pairing state is most consistent with our experiments (see Table\,\ref{T2}). The reason is that for the $B_{3u}+iA_u$ state with $|{\bm d}_{B3u}| \approx |{\bm d}_{Au}|$ conditions (Fig.\,\ref{F1}e), multiple point nodes can exist near the $k_y$- and $k_z$-axes, leading to larger excitations along the $b$- and $c$-axes than the single order parameter cases. Thus, the quicker decrease of our $\rho_{{\rm s,}b}$ and $\rho_{{\rm s,}c}$ data than the theoretical $\bm{j}_s \parallel \bm{I}$ curve is consistent with the $B_{3u}+iA_u$ pairing state. This state can be supported by a recent theoretical study based on the periodic Anderson model, which suggests almost equally stable $B_{3u}$ and $A_u$ states at ambient pressure\,\cite{Ishizuka2021}. We note that the chiral $B_{3u}+iA_u$ state is non-unitary with finite $\bm{d}\times\bm{d}^*$ (see Supplementary Information IV), and for the system close to a ferromagnetic quantum critical point, theory shows that such a non-unitary complex order parameter may become stable\,\cite{Nevidomskyy_arxiv}. Moreover, recent studies of NMR Knight shift\,\cite{
Nakamine2021} suggest finite ${\hat y}$ and ${\hat z}$ components of ${\bm d}$, which is consistent with the $B_{3u}+iA_u$ state.

Next we discuss the small exponent value $n_c=1.45$ and 1.84 for crystals \#A1 and \#B1, respectively, for $c$-axis penetration depth (Figs.\,\ref{F4}a,b). In the line node case, the exponent value larger than the clean limit $n=1$ and smaller than $n=2$ can be interpreted as a consequence of nonmagnetic impurity scatterings\,\cite{Hirschfeld1993}, quantum criticality\,\cite{Hashimoto2013
}, or non-local effects\,\cite{Kosztin1997}. However, these possibilities can be excluded in the case of UTe$_2$ (see Supplementary Information III). Moreover, since the existence of line nodes is not allowed in the odd-parity order parameters in UTe$_2$, these mechanisms cannot be applied to UTe$_2$. We also note that, in the presence of point nodes in the gap structure, the impurity scattering can affect only the amplitude of the low-energy excitations, but does not change the exponent value of $\Delta \lambda (T) \propto T^2${\,}\cite{Bae2021,Gross1986}. Here we propose an interference effect of the two point nodes located closely. In the $B_{3u}+iA_u$ state, the main contributions to $\Delta \lambda_c(T)/\lambda_c(0)$ come from two pairs of two point nodes near the north and south poles along the $k_z$-axis (Fig.\,\ref{F1}e). When the distance of the two nodes near the pole gets sufficiently short, the low-energy excitations can no longer be treated as a sum of the contributions from two independent point nodes, and this interference effect can lead to an exponent value less than 2. We have confirmed that a simple model based on the $B_{3u}+iA_u$ state can indeed lead to an exponent near the experimental $n_c$ value when the two point nodes are sufficiently close to each other (see Supplementary Figs.\,S12 and S13). Moreover, we note that larger exponent values in crystal \#B1 compared to crystal \#A1 can be explained by varying a parameter in the $A_u$ component (for more details, see Supplementary Information IX). 
However, for detailed understanding of the microscopic mechanism of the sample-dependent anisotropy in the order parameter, further studies are desired.

We should mention that theoretical and ARPES studies both suggest the presence of a heavy three-dimensional Fermi surface (FS) around the $Z$ point, in addition to quasi-two-dimensional FSs mainly formed by U $d$ orbitals and Te $p$ orbitals\,\cite{Miao2020,Ishizuka2019,Ishizuka2021,Shishidou2021}. The presence of the three-dimensional FS is also supported by the relatively isotropic transport properties\,\cite{Eo_arxiv}. These indicate that there are FSs near $k_x$-, $k_y$-, and $k_z$-axes, which ensures the validity of our discussions on the nodal structure.

Finally, we note that our conclusion of the $B_{3u}+iA_u$ pairing state is apparently different from the recent report of field angle-resolved specific heat measurements suggesting the point nodes only on the $k_x$-axis\,\cite{Kittaka2020}. However, our anisotropic measurements in the Meissner state probe the ground state in the zero-field limit, while the application of strong magnetic field can change the superconducting symmetry\,\cite{Ishizuka2019,Nevidomskyy_arxiv,
Ishizuka2021,Shishidou2021}. How the $B_{3u}+iA_u$ state found here changes as a function of field deserves further investigations.


To sum up, from the anisotropic penetration depth measurements, 
we have found larger excitations along the $b$- and $c$-axes than along the $a$-axis. 
This rules out any of single-component odd-parity states, and thus indicates a multi-component order parameter. 
The analysis reveals the presence of multiple point nodes near the $k_y$- and $k_z$-axes, which is most consistent with the chiral $B_{3u}+iA_u$ superconducting state in UTe$_2$. The presence of TRSB components splits the point node to multiple point nodes away from high-symmetry axes, and in analogy to topological Weyl points in Weyl semimetals, these nodes are expected to create surface arcs of zero-energy Majorana quasiparticles states. Thus UTe$_2$ is an ideal platform to investigate chiral superconductivity and its related topological physics. In particular, the positions of multiple point nodes (Weyl points) in the bulk studied here are fundamentally important to determine the topological properties of surface states.   

\section*{Methods}
Single crystals of UTe$_2$ were grown by the chemical vapor transport method with iodine as the transport agent. The details of the characterizations of the crystals are shown in Supplementary Information V. 

To obtain anisotropic components of penetration depth $\Delta \lambda_a (T)$, $\Delta \lambda_b (T)$, and $\Delta \lambda_c (T)$ data separately, we have performed high-precision measurements of ac magnetic susceptibility shift $\Delta \chi (T)\equiv \chi (T)-\chi (0)$ using a tunnel diode oscillator technique operated at 13.8\,MHz with weak ac magnetic field $H_\omega$ along the three crystallographic axes\,\cite{Prozorov_review,Prozorov_arxiv
}. The ac magnetic field $H_\omega$ induced by the coil of the oscillator is the order of $\mu$T, which is much lower than the lower critical field of the order of mT in UTe$_2$\,\cite{Paulsen_arxiv}. In this technique, the frequency shift of the oscillator $\Delta f(T) \equiv f(T) - f(0)$ is proportional to $\Delta \chi (T)$. 

Specific heat capacity was measured by a long relaxation method in a $^3$He cryostat, where a Cernox resistor is used as a thermometer, a heater and a sample stage. The bare chip is suspended from the cold stage in order that it has weak thermal link to the cold stage, and electrical connection for the sensor reading. The samples are mounted on the bare chip using Apiezon N grease. The specific heat of the crystals is obtained by subtracting the heat capacity of bare chip and grease from the total data.


\section*{Acknowledgments}
We thank S. Fujimoto, J. Ishizuka, T. Matsushita and Y. Yanase for fruitful discussions, and N. Abe, Y. Tokunaga, and T. Arima for technical supports. This work was supported by Grants-in-Aid for Scientific Research (KAKENHI) (Nos. JP21J10737, JP21H01793, JP19H00649, JP18H05227, JP16KK0106), Grant-in-Aid for Scientific Research on innovative areas ``Quantum Liquid Crystals" (No.\ JP19H05824), Grant-in-Aid for Scientific Research for Transformative Research Areas (A) ``Condensed Conjugation'' (No.\ JP20H05869) from Japan Society for the Promotion of Science (JSPS), and CREST (No.\ JPMJCR19T5) from Japan Science and Technology (JST). 

\section*{Data Availability}
The data that support the findings of this study are available from the corresponding
authors upon reasonable request.

\clearpage

\begin{table}[t]
    \centering
    \caption{\bf Basis functions, nodal types, and temperature dependence of the magnetic penetration depth for odd-parity order parameters in the point group $D_{2h}$.}
    \label{T1}
\begin{tabular*}{1.0\columnwidth}{@{\extracolsep{\fill}}cccc}
    \hline\hline
    IR & Basis functions & Nodes & $\Delta \lambda (T)$ \\ \hline
    $A_u$ & $k_x \hat{x}, k_y \hat{y}, k_z \hat{z}$ & None & $\Delta \lambda_{a,b,c} \propto {\rm exp} (-|{\bm d}|/T)$ \\
    $B_{1u}$ & $k_y \hat{x}, k_x \hat{y}, k_x k_y k_z \hat{z}$ & Point ($k_z$) & $\Delta \lambda_{a,b} \propto T^4$, $\Delta \lambda_{c} \propto T^2$ \\
    $B_{2u}$ & $k_z \hat{x}, k_x k_y k_z \hat{y}, k_x \hat{z}$ & Point ($k_y$) & $\Delta \lambda_{c,a} \propto T^4$, $\Delta \lambda_{b} \propto T^2$ \\
    $B_{3u}$ & $k_x k_y k_z \hat{x}, k_z \hat{y}, k_y \hat{z}$ & Point ($k_x$) & $\Delta \lambda_{b,c} \propto T^4$, $\Delta \lambda_{a} \propto T^2$ \\
    \hline\hline
\end{tabular*}
    \\[5mm]
    \caption{\bf Expected positions of point nodes in chiral superconducting states with the order parameter ${\bm d}={\bm d}_1+i{\bm d}_2$, where ${\bm d}_1$ and ${\bm d}_2$ are classified into different IRs.}
    \label{T2}
\begin{tabular*}{1.0\columnwidth}{@{\extracolsep{\fill}}cc|ccc}
    \hline\hline
    \multicolumn{2}{c|}{IR} & \multicolumn{3}{c}{Positions of point nodes} \\
    ${\bm d_1}$ & ${\bm d_2}$ & $|{\bm d_1}| \ll |{\bm d_2}|$ & $|{\bm d_1}| \approx |{\bm d_2}|$ & $|{\bm d_1}| \gg |{\bm d_2}|$ \\ \hline
    $B_{1u}$ & $B_{2u}$ & near $k_y$-axis & near $k_x$-axis & near $k_z$-axis \\
    $B_{2u}$ & $B_{3u}$ & near $k_x$-axis & near $k_z$-axis & near $k_y$-axis \\
    $B_{3u}$ & $B_{1u}$ & near $k_z$-axis & near $k_y$-axis & near $k_x$-axis \\
    $B_{1u}$ & $A_u$ & None & near $k_x$- and/or $k_y$-axes & near $k_z$-axis \\
    $B_{2u}$ & $A_u$ & None & near $k_z$- and/or $k_x$-axes & near $k_y$-axis \\
    $B_{3u}$ & $A_u$ & None & near $k_y$- and/or $k_z$-axes & near $k_x$-axis \\
    \hline\hline
\end{tabular*}
\end{table}
    
\begin{figure}[tbp]
    \includegraphics[width=\linewidth]{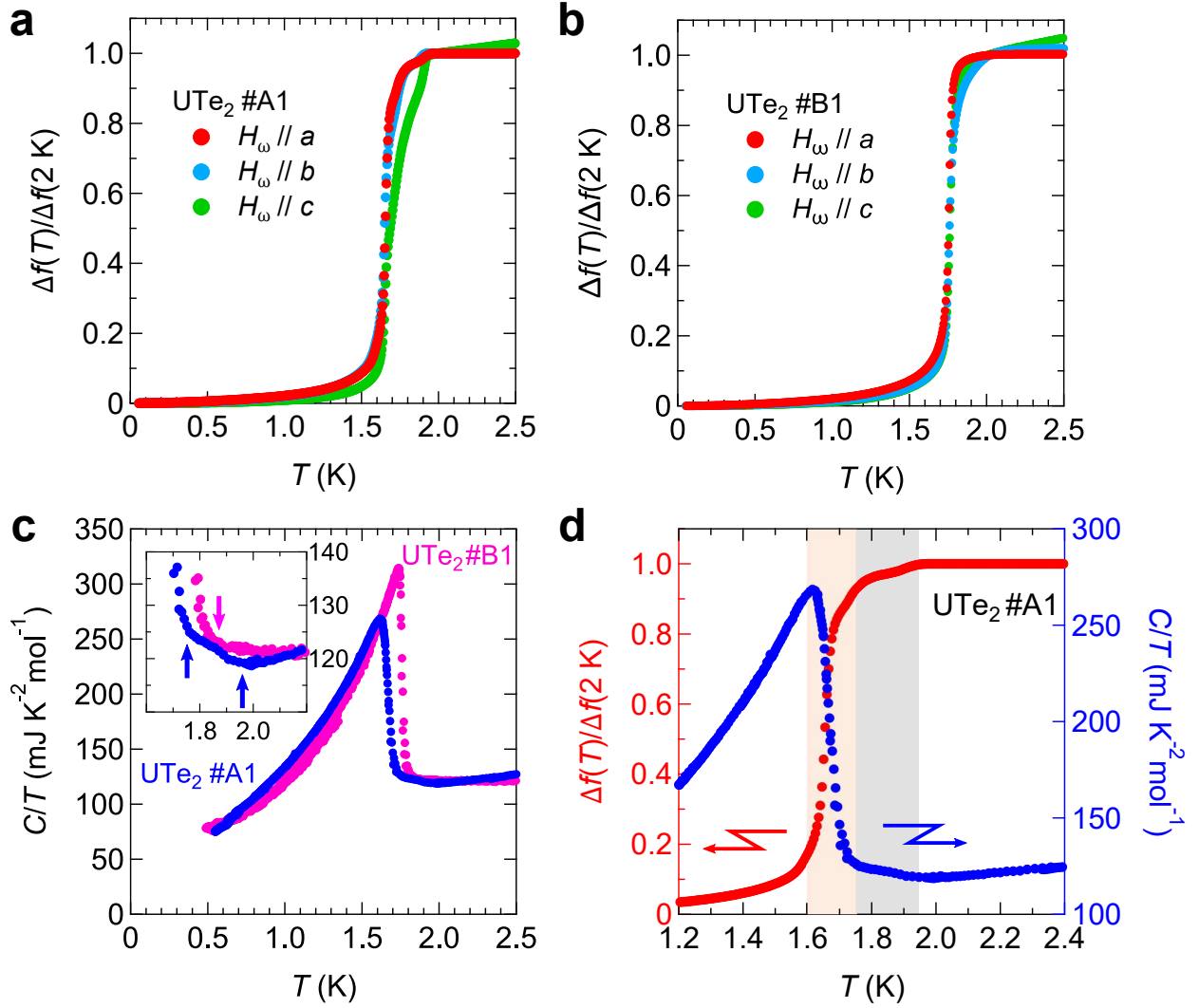}
    \caption{{\bf Superconducting transitions in UTe$_2$.} {\bf a-b}, Overall temperature dependence of the frequency shift $\Delta f$ in UTe$_2$ for crystal \#A1 ({\bf a}) and \#B1 ({\bf b}). All data are normalized by the values at 2\,K.  {\bf c}, Temperature dependence of the specific heat $C$ divided by $T$ in crystals \#A1 and \#B1 measured at zero field. 
    {\bf d}, Enlarged views of $\Delta f(T)$ for the $a$-axis field (red) and $C(T)/T$ at zero field (blue) around the superconducting transition in crystal \#A1. Two superconducting transitions at $T_c=1.65$\,K and $T_c\sim 1.9$\,K are hatched by red and gray, respectively.}
    \label{F2}
\end{figure}

\clearpage

\setcounter{figure}{0}
\setcounter{table}{0}
\renewcommand{\tablename}{Table S$\!\!$}
\renewcommand{\thetable}{\arabic{table}}
\renewcommand{\figurename}{FIG. S$\!\!$}
\renewcommand*{\citenumfont}[1]{S#1}
\renewcommand*{\bibnumfmt}[1]{[S#1]}
\renewcommand{\theequation}{S\arabic{equation}}
\newcommand{\vect}[1]
{{\mbox{\boldmath $#1$}}}


\begin{center}
{\Large \bf Supplementary Information}
\end{center}

\section{Positions of point nodes in chiral spin-triplet superconducting states}
Here, we discuss the positions of point nodes in chiral superconducting states whose order parameters consist of two components represented by two odd-parity irreducible representations (IRs). First, we consider the $B_{3u}+iB_{1u}$ state. From the basis functions summarized in Table\,I in the main text, the vector order parameter ${\bm d}$ is written by
\begin{equation}
{\bm d} = {\bm d}_{B3u} + i{\bm d}_{B1u} = \left(
\begin{array}{c}
c_1 k_x k_y k_z \\
c_2 k_z \\
c_3 k_y
\end{array}
\right) + i \left(
\begin{array}{c}
c_4 k_y \\
c_5 k_x \\
c_6 k_x k_y k_z
\end{array}
\right),
\label{B3u_iB1u}
\end{equation}
where $c_1$ to $c_6$ are real coefficients. We note that higher-order terms are neglected, which do not change our discussions below as long as considering a spherical Fermi surface (FS) around the $\Gamma$ point. In the chiral spin-triplet state, the positions of point nodes in the order parameter ${\bm d} = {\bm d}_1 +i{\bm d}_2$ are derived by the two conditions\,\cite{Hayes_science}
\begin{numcases}{}
\label{node_1}
{\bm d}_1 \cdot {\bm d}_2 =0 & \\
\label{node_2}
|{\bm d}_1| = |{\bm d}_2|. &
\end{numcases}
By applying the above conditions to the order parameter in Eq.\,\ref{B3u_iB1u}, the nodes appear at
\begin{equation}
\left\{
\begin{aligned}
& k_z=0 \\
& (c_3^2-c_4^2)k_y^2=c_5^2 k_x^2
\end{aligned}
\right. \qquad {\rm or} \qquad \left\{
\begin{aligned}
& k_x=0 \\
& (c_4^2-c_3^2)k_y^2=c_2^2 k_z^2
\end{aligned}
\right..
\end{equation}
Apparently, when $|c_3|>|c_4|$ ($|c_3|<|c_4|$), the first (second) condition induces point nodes, and, with the spherical FS, four point nodes appear on the $k_z=0$ ($k_x=0$) plane. Furthermore, qualitatively speaking, when $|{\bm d}_{B1u}|$ is much smaller than $|{\bm d}_{B3u}|$ ($|c_3| \gg |c_4|, |c_5|$), point nodes appear near the $k_x$-axis (Fig.\,1c), and when $|{\bm d}_{B1u}| \approx |{\bm d}_{B3u}|$ ($|c_3| \approx |c_4|$), point nodes appear near the $k_y$-axis.

Next, we consider the $B_{3u}+iA_u$ state. The order parameter ${\bm d}$ can be written as
\begin{equation}
{\bm d} = {\bm d}_{B3u} + i{\bm d}_{Au} = \left(
\begin{array}{c}
c_1 k_x k_y k_z \\
c_2 k_z \\
c_3 k_y
\end{array}
\right) + i \left(
\begin{array}{c}
c_4 k_x \\
c_5 k_y \\
c_6 k_z
\end{array}
\right).
\label{B3u_iAu}
\end{equation}
By applying Eqs.\,\ref{node_1} and \ref{node_2} to Eq.\,\ref{B3u_iAu}, we obtain the conditions
\begin{equation}
\left\{
\begin{aligned}
& k_y=0 \\
& (c_2^2-c_6^2)k_z^2=c_4^2 k_x^2
\end{aligned}
\right. \qquad {\rm and} \qquad \left\{
\begin{aligned}
& k_z=0 \\
& (c_3^2-c_5^2)k_y^2=c_4^2 k_x^2
\end{aligned}
\right..
\end{equation}
In this case, several nodal structures become possible. First, when $|c_2|>|c_6|$ and $|c_3|>|c_5|$, four point nodes appear on both $k_y=0$ and $k_z=0$ planes, and thus eight point nodes appear in total on the spherical FS. Next, when $|c_2|>|c_6|$ and $|c_3|<|c_5|$ ($|c_2|<|c_6|$ and $|c_3|>|c_5|$), four point nodes appear on the $k_y=0$ ($k_z=0$) plane. Finally, when $|c_2|<|c_6|$ and $|c_3|<|c_5|$, the superconducting state is fully gapped. Especially, when $|{\bm d}_{Au}|$ is much smaller than $|{\bm d}_{B3u}|$ ($|c_2| \gg |c_4|, |c_6|$ and $|c_3| \gg |c_4|, |c_5|$), point nodes are near the $k_x$-axis (Fig.\,1d), and when $|{\bm d}_{Au}| \approx |{\bm d}_{B3u}|$ ($|c_2| \approx |c_6|$ and $|c_3| \approx |c_5|$), point nodes may appear near the $k_y$- and $k_z$-axes (Fig.\,1e).
\\

\section{Gap structures with two odd-parity order parameters preserving time-reversal symmetry}

\begin{figure}[tbp]
\includegraphics[width=\linewidth]{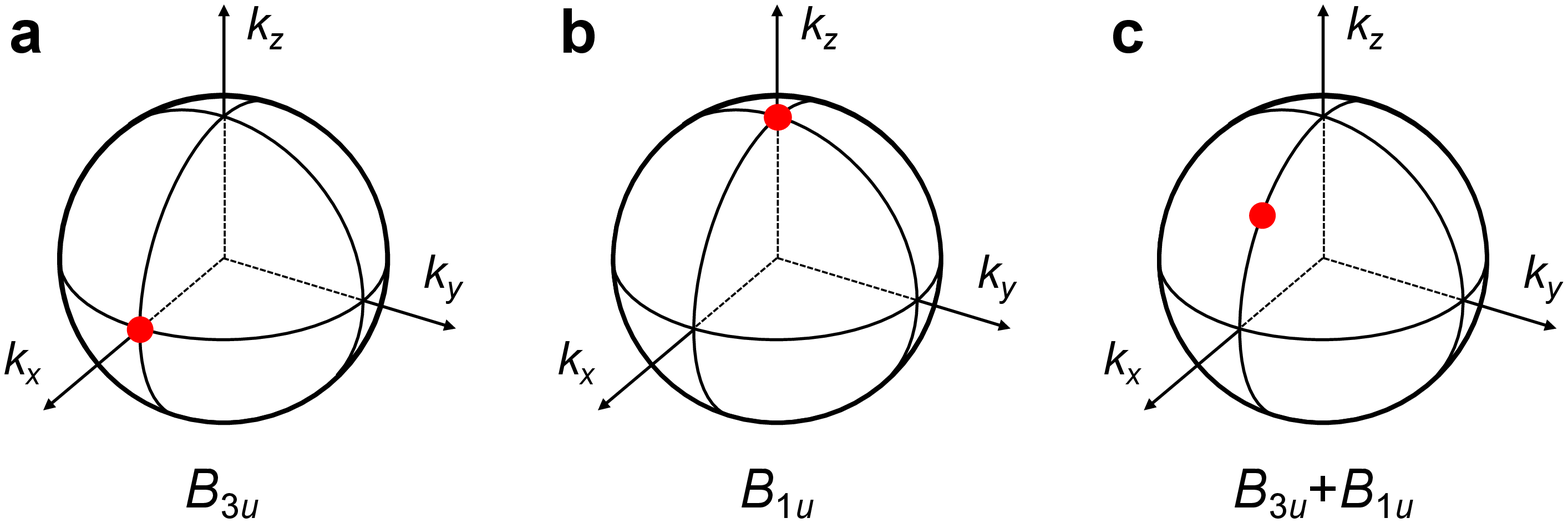}
\caption{{\bf Nodal positions induced by $B_{3u}$ and $B_{1u}$ states preserving time-reversal symmetry.} {\bf a-c,} Positions of the point nodes for the $B_{3u}$ ({\bf a}), $B_{1u}$ ({\bf b}),  and $B_{3u}+B_{1u}$ states ({\bf c}).}
\label{F_B3u_B1u}
\end{figure}

In this section, we discuss the gap structures of the order parameter ${\bm d} = {\bm d}_1 + {\bm d}_2$ preserving time-reversal symmetry, where ${\bm d}_1$ and ${\bm d}_2$ belong to different IRs. When ${\bm d}_1$ and ${\bm d}_2$ are represented by $B_{3u}$ and $B_{1u}$ symmetries, respectively, ${\bm d}$ can be written as
\begin{equation}
{\bm d} = {\bm d}_{B3u} + {\bm d}_{B1u} = \left(
\begin{array}{c}
c_1 k_x k_y k_z + c_4 k_y \\
c_2 k_z + c_5 k_x \\
c_3 k_y + c_6 k_x k_y k_z
\end{array}
\right).
\label{B3u_B1u}
\end{equation}
In this case, the positions of nodes are determined from the condition $|{\bm d}|=0$. Therefore, point nodes satisfy the relations
\begin{equation}
\left\{
\begin{aligned}
& k_y=0 \\
& c_2k_z=-c_5 k_x
\end{aligned}
\right.,
\end{equation}
meaning that two point nodes appear on the $k_y = 0$ plane but off the high-symmetry axes (Fig.\,S\ref{F_B3u_B1u}c). As mentioned in the main text, because the original point nodes of the $B_{3u}$ state is not split by adding the $B_{1u}$ component, the $B_{3u}+B_{1u}$ state cannot account for our experimental results.

In the $B_{3u}+A_u$ state, ${\bm d}$ can be represented by
\begin{equation}
{\bm d} = {\bm d}_{B3u} + {\bm d}_{Au} = \left(
\begin{array}{c}
c_1 k_x k_y k_z + c_4 k_x \\
c_2 k_z + c_5 k_y \\
c_3 k_y + c_6 k_z
\end{array}
\right).
\label{B3u_Au}
\end{equation}
For the emergence of the point nodes satisfying $|{\bm d}|=0$, the coefficients need to satisfy a specific condition $c_2/c_5=c_3/c_6$. When this condition is valid, the nodes appear in the points represented by
\begin{equation}
\left\{
\begin{aligned}
& k_x=0 \\
& c_2k_z=-c_5 k_y
\end{aligned}
\right.,
\end{equation}
and non-split nodes are located on the $k_x=0$ plane. Therefore, a fine-tuning of the coefficients is required for the emergence of nodes, and even if the coefficients satisfy the specific condition, the $B_{3u}+A_u$ state is inconsistent with our experimental results.
\\

\section{Gap structures in the superconducting states with even-parity order parameters}

The even-parity IRs of the point group of $D_{2h}$ are listed in Table\,S\ref{ST1}. In the $A_{1g}$ state, a fully-gapped state is realized, which is clearly inconsistent with our experimental results. On the other hand, the $B_{1g}$, $B_{2g}$, and $B_{3g}$ states lead to gap structures with line nodes. As mentioned in the main text, a line nodal structure can induce the exponent value $1\leq n\leq 2$ of the power-law fitting $\Delta \lambda \propto T^n$ because of nonmagnetic impurity scatterings\,\cite{Hirschfeld1993}, quantum criticality\,\cite{Hashimoto2013,Mizukami2016}, and non-local effects\,\cite{Kosztin1997}. In the case of UTe$_2$, although ferromagnetic quantum criticality is expected\,\cite{Ran2019_SI}, the mass renormalization cannot account for the experimental exponents $n_a \approx n_b \approx 2$. Besides, the non-local effects are expected to be too weak due to $\xi/\lambda(0) \sim 0.01$. For the estimation of the strength of impurity scatterings, a pair breaking parameter $g\equiv \hbar / \tau_{\rm imp}k_{\rm B} T_{c0}$ is usually considered\,\cite{Prozorov2014,Takenaka2017}, where $\hbar$ and $k_{\rm B}$ are Dirac and Boltzmann constants, respectively, $\tau_{\rm imp}$ is impurity scattering time, and $T_{\rm c0}$ is the superconducting transition temperature in the clean limit. $\tau_{\rm imp}$ can be calculated from $\tau_{\rm imp} = \mu_0 \lambda^2 (0)/\rho_0$, where $\rho_0$ is the residual resistivity. By substituting the values $T_{\rm c0}=1.65$\,K, $\lambda (0)=1$\,$\mu$m, and $\rho_0 = 30$\,$\mu \Omega$cm (Fig.\,S\ref{Specific_Resis}c), we obtain $g=1.1$. Compared with the $d$-wave superconductor CeCoIn$_5$, this $g$ value seems to be too small to induce the exponent value about 2\,\cite{Takenaka2017,Kim2015}, but in the case of YBCO, $\Delta\lambda(T)\propto T^2$ has been observed even for $g < 1.1$ \cite{Bonn1994}. Therefore, the pair breaking parameter $g = 1.1$ in UTe$_2$ itself does not exclude the possibility that $\Delta\lambda(T)\propto T^2$ is caused by the line node gap structure with impurity scatterings. In order to discuss the impurity effect in UTe$_2$ quantitatively, we compare the resistivities of crystals \#A1 and \#B1 derived from the skin depth (see section VI). From the $\Delta\chi_c$ data in Fig.\,S\ref{chi}, the skin depths of crystals \#A1 and \#B1 are estimated to be $\sim 75\,\mu$m and $\sim 50\,\mu$m, respectively, which correspond to $\rho \sim 30\,\mu\Omega$cm and $\sim 14\,\mu\Omega$cm, respectively. Thus, crystal \#B1 is a cleaner sample than crystal \#A1. If $\Delta\lambda(T)\propto T^2$ comes from a line node gap structure with impurity scatterings, the exponent $n_c$ of crystal \#B1 should be smaller than that of crystal \#A1, which is opposite to the experimental results. We also note that several groups have reported universally $\Delta\lambda_{ab}\propto T^2$ in different samples with different $T_c$s, indicating that $\Delta\lambda_{ab}\propto T^2$ is an intrinsic property of UTe$_2$. Thus, our discussions above suggest that the line nodal gap structure with impurity scatterings cannot account for our experimental results (here we note that, in the presence of point nodes in the gap structure, the impurity scattering can affect only the amplitude of the low-energy excitations, but does not change the exponent value of $\Delta \lambda (T) \propto T^2${\,}[9, 26]). Moreover, in the case of UTe$_2$, several physical properties, such as Knight shift, upper critical field, and reentrant superconductivity, strongly suggest the spin-triplet superconducting state. In spin-triplet states with a symmorphic crystal structure, the existence of line nodes is prohibited from Blount's theorem, which is obvious in the irreducible representations of the point group  (see Table I in the main text).

Next, we discuss mixed-parity states consisting of even-parity order parameter $\psi({\bm k})$ and odd-parity order parameter ${\bm d} ({\bm k})$. We note that the mixed-parity state in UTe$_2$ requires the inversion symmetry breaking which is proposed in Ref.\,\cite{Ishizuka2021_SI} under hydrostatic pressure. When the time-reversal symmetry is broken, because the gap size can be written as $\Delta_{\rm gap} ({\bm k}) = \sqrt{|\psi({\bm k})|^2+|{\bm d}({\bm k})|^2}$, the superconducting state is fully-gapped or has point nodes on the high-symmetry axes. These states are inconsistent with our experimental results.

\begin{figure}[tbp]
\includegraphics[width=\linewidth]{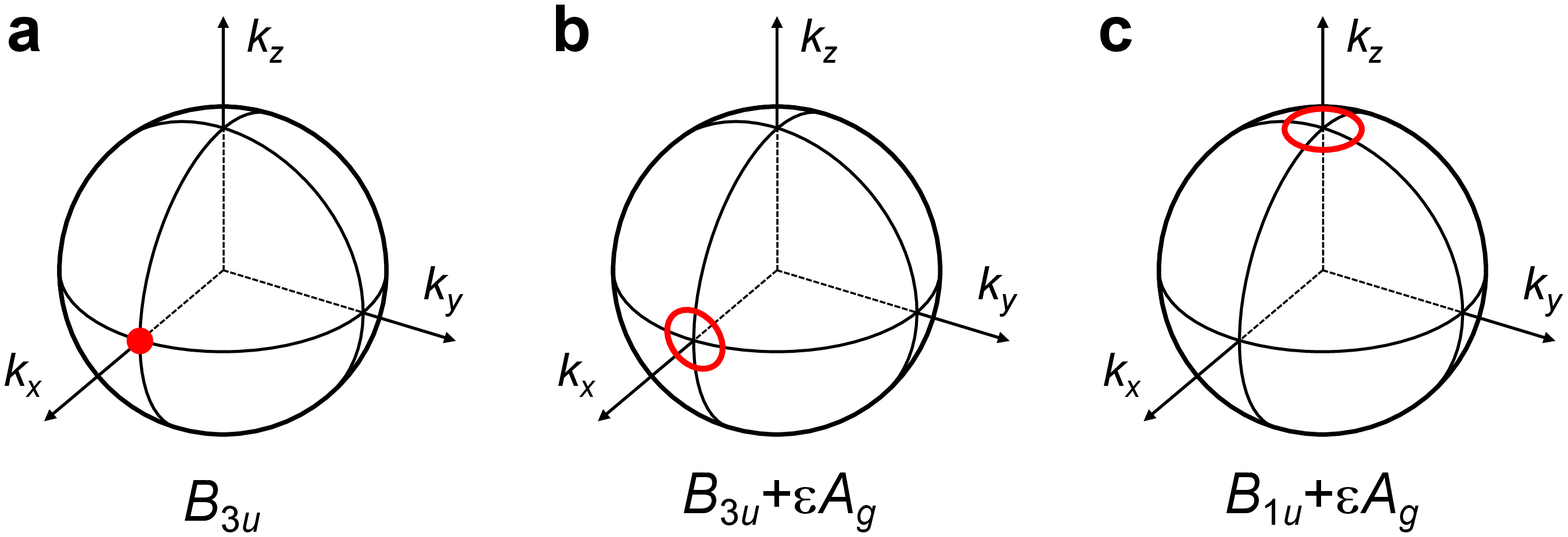}
\caption{{\bf Nodal positions induced by $B_{3u}$ and $A_u$ states preserving time-reversal symmetry.} {\bf a-c,} Positions of the nodes for  the $B_{3u}$ ({\bf a}), $B_{3u}+\varepsilon A_g$ ({\bf b}), and $B_{1u}+\varepsilon A_g$ states ({\bf c}), where $\varepsilon$ is a small real number.}
\label{B3u_Ag}
\end{figure}

When the time-reversal symmetry is preserved, on the other hand, the spin split FSs have a different gap size $\Delta_{\rm gap}^{\pm} ({\bm k}) = \psi({\bm k}) \pm |{\bm d}({\bm k})|$, leading to various gap structures. In this study, we focus on the mixed-parity states of the $B_{1u}$ or $B_{3u}$ and ${A_g}$ states, the latter of which is indeed proposed in Ref.\,\cite{Ishizuka2021_SI}. When the small $A_g$ component is admixed with the $B_{3u}$ component, the original point nodes on the $k_x$-axis spread into the nodal rings around the $k_x$-axis (Fig.\,S\ref{B3u_Ag}b). This nodal structure is inconsistent with our experiments. On the other hand, since the existence of the small nodal rings can induce an exponent value of $1\leq n \leq 2$, the $B_{1u}+A_g$ state with the nodal rings around $k_z$-axis may partly be consistent with our experiments. However, as seen in Fig.\,S\ref{B3u_Ag}c, this state cannot explain the large anisotropy between $\Delta \lambda_a (T)$ and $\Delta \lambda_b (T)$. Thus, we can rule out these mixed-parity superconducting states from the experimental results.
\\

\section{Non-unitary pairing states}

Non-unitary pairing states are defined by ${\bm q} ({\bm k}) \equiv i{\bm d}({\bm k}) \times {\bm d}^* ({\bm k}) \neq 0$\,\cite{Sigrist_Ueda}. In the $B_{3u}+iA_u$ state, ${\bm q({\bm k})}$ can be written as
\begin{equation}
{\bm q}({\bm k}) =i {\bm d}({\bm k}) \times {\bm d}^* ({\bm k}) = -2i {\bm d}_{B3u}({\bm k}) \times {\bm d}_{Au} ({\bm k}) = -2i \left(
\begin{array}{c}
c_2 c_6 k_z^2 - c_3 c_5 k_y^2 \\
k_x k_y (c_3 c_4 - c_1 c_6 k_z^2) \\
k_z k_x (-c_2 c_4 + c_1 c_5 k_y^2)
\end{array}
\right).
\label{non_unitary}
\end{equation}
Therefore, the $B_{3u}+iA_u$ state is non-unitary. Physically, this ${\bm q}({\bm k})$ corresponds to an orbital moment of a Cooper pair, and in the $B_{3u}+iA_u$ case, the average of ${\bm q}({\bm k})$ in the FSs $\left< {\bm q}({\bm k}) \right>_{\rm FS}$ is parallel to $\hat{x}$ which is the magnetic easy axis in the normal state\,\cite{Ran2019_SI}. Usually, the non-unitary state is energetically unstable except for the ferromagnetic superconductors\,\cite{Aoki_review_SI}. However, a previous theoretical study\,\cite{Nevidomskyy_arxiv_SI} suggests that, when a material is close to a ferromagnetic quantum critical point as in the case of UTe$_2$\,\cite{Ran2019_SI}, the non-unitary pairing states can be stable even in a paramagnetic state.
\\

\section{Sample characterization}

\begin{figure}[tbp]
\includegraphics[width=\linewidth]{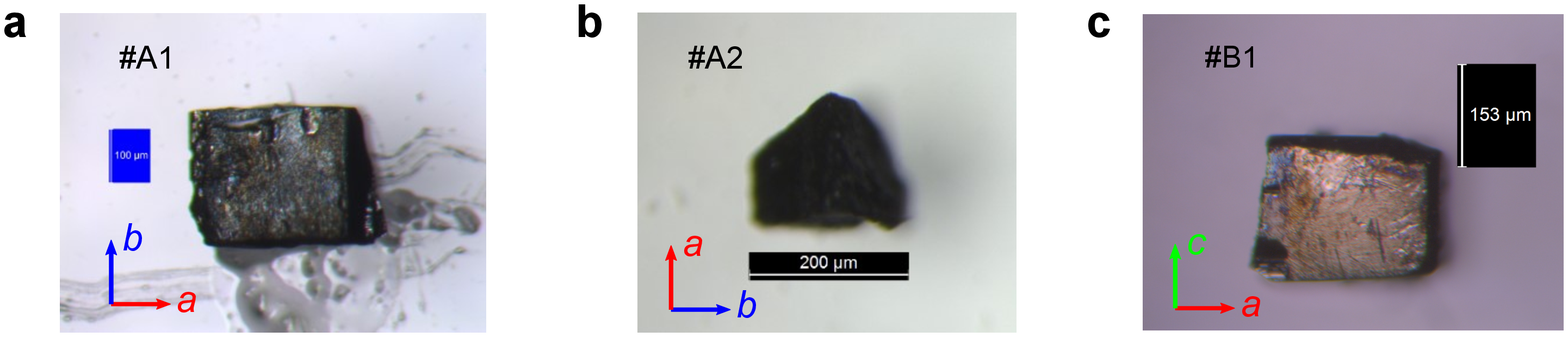}
\caption{{\bf Geometry of the UTe$_2$ single crystals.} {\bf a-c,} Pictures of single crystals \#A1 ({\bf a}), \#A2 ({\bf b}), and \#B1 ({\bf c}).}
\label{Single_crystals}
\end{figure}

Single crystals of UTe$_2$ were grown by the chemical vapor transport method with iodine as the transport agent. The shapes of crystals \#A1, \#A2, and \#B1 are depicted in Figs.\,S\ref{Single_crystals}a-c, respectively. Crystals \#A1 and \#B1 have a cuboid shape with dimensions $300 \times 250 \times 35$\,$\mu$m$^3$ and $260 \times 155 \times 210$\,$\mu$m$^3$, respectively, while crystal \#A2 has a more complicated shape. As described later, the $\Delta \lambda_i (T)$ analysis is based on a cuboid shape, we focus on the data of crystals \#A1 and \#B1 in the $\Delta \lambda_i (T)$ discussions below. We note that the data of $\Delta f(T)$ providing the similar exponent values for all the crystals in the power-law fittings, $\Delta f(T) \propto T^n$, imply that qualitatively similar $\Delta \lambda_i (T)$ results with crystals \#A1 and \#B1 are also expected in crystal \#A2.

\begin{figure}[tbp]
\includegraphics[width=\linewidth]{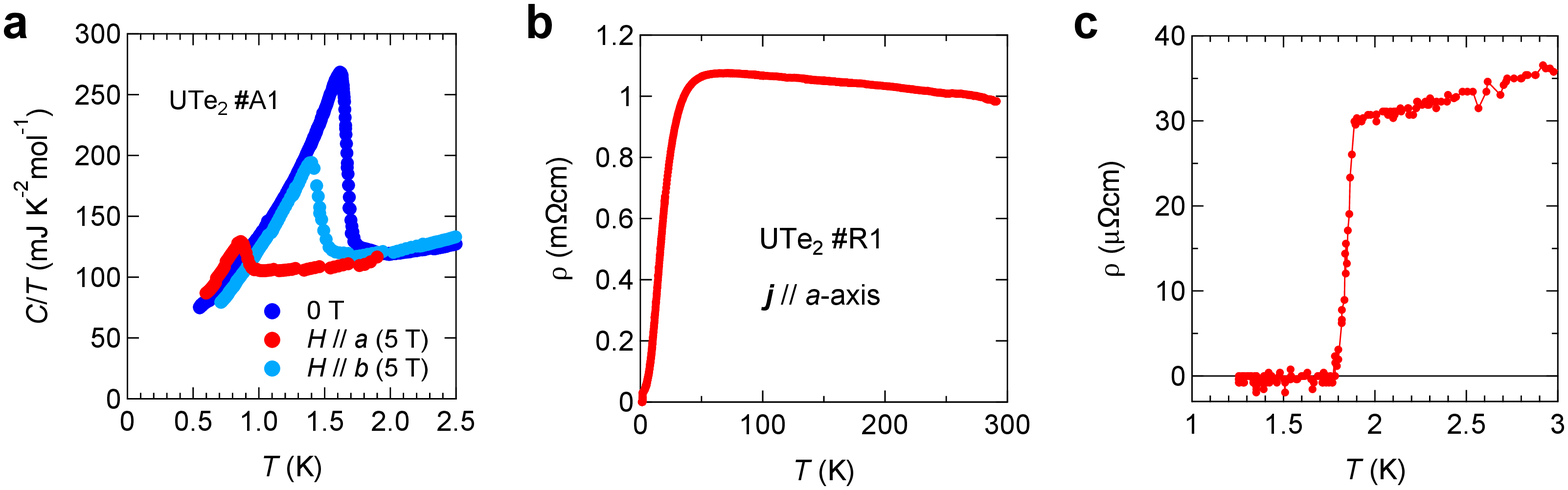}
\caption{{\bf Specific heat and resistivity.} {\bf a,} Specific heat of crystal \#A1 at zero field and under 5\,T along the $a$- and $b$-axes. {\bf b,c,} The resistivity of crystal \#R1 with the current along the $a$-axis in whole $T$ range ({\bf b}) and at low $T$ ({\bf c}).}
\label{Specific_Resis}
\end{figure}

Figure\,S\ref{Specific_Resis}a shows specific heat of crystal \#A1 at zero field and under $\mu_0 H =5$\,T along the $a$- and $b$-axes. The difference of $T_c$ between $H \parallel a$ and $H \parallel b$ suggests a large anisotropy of the upper critical field as already demonstrated in previous studies. Figure\,S\ref{Specific_Resis}b depicts the resistivity with the current direction of $a$-axis in crystal \#R1 picked up from the same ampoule as crystals \#A1 and \#A2. The resistivity shows maximum around 50\,K below which Kondo hybridization yields coherent electronic states. The resistivity at low $T$ is shown in Fig.\,S\ref{Specific_Resis}c. The onset of the superconducting transition temperature $T_c^{\rm on}$ is about 1.89\,K, and the resistivity gets zero at $T_c^{\rm zero}\approx 1.77$\,K. Compared with $T_c \approx 1.65$\,K obtained from a large change of $\Delta f(T)$ and $C/T$, $T_c^{\rm on}$ and $T_c^{\rm zero}$ are relatively high, and $T_c^{\rm on}$ is close to the onset of the small diamagnetic signal in $\Delta f(T)$ and the small jump of $C/T$ (Fig.\,2d). This indicates that a superconducting path exists between the electrodes even above $T_c\approx 1.65$\,K.
\\


\section{Derivation of anisotropic penetration depth from frequency shift}

\begin{figure}[tbp]
\includegraphics[width=0.4\linewidth]{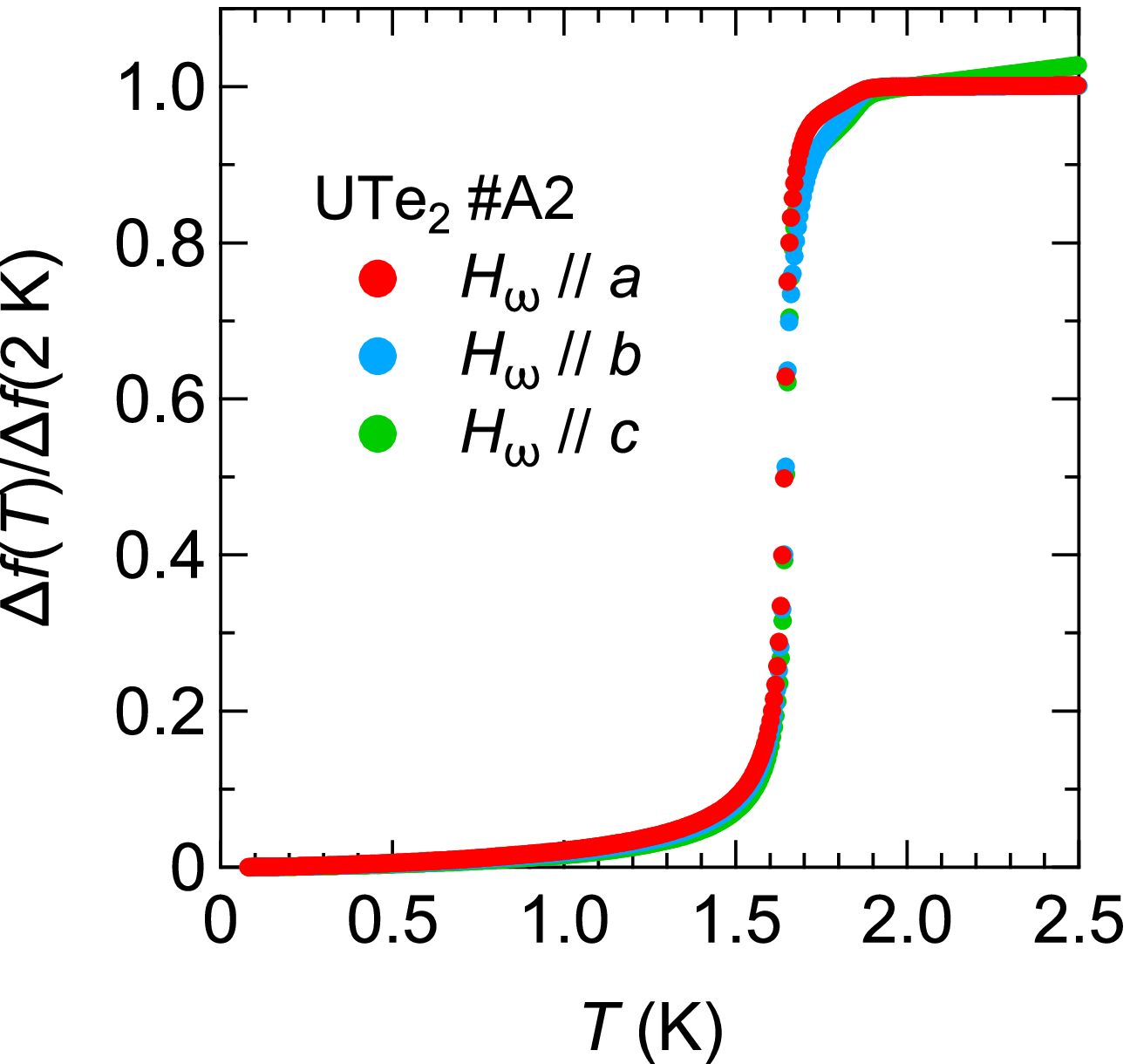}
\caption{{\bf Frequency shift for crystal \#A2.} Overall temperature dependence of the frequency shift normalized by the value at 2\,K with the magnetic field along each crystallographic axis.}
\label{sample2}
\end{figure}

\begin{figure}[tbp]
\includegraphics[width=\linewidth]{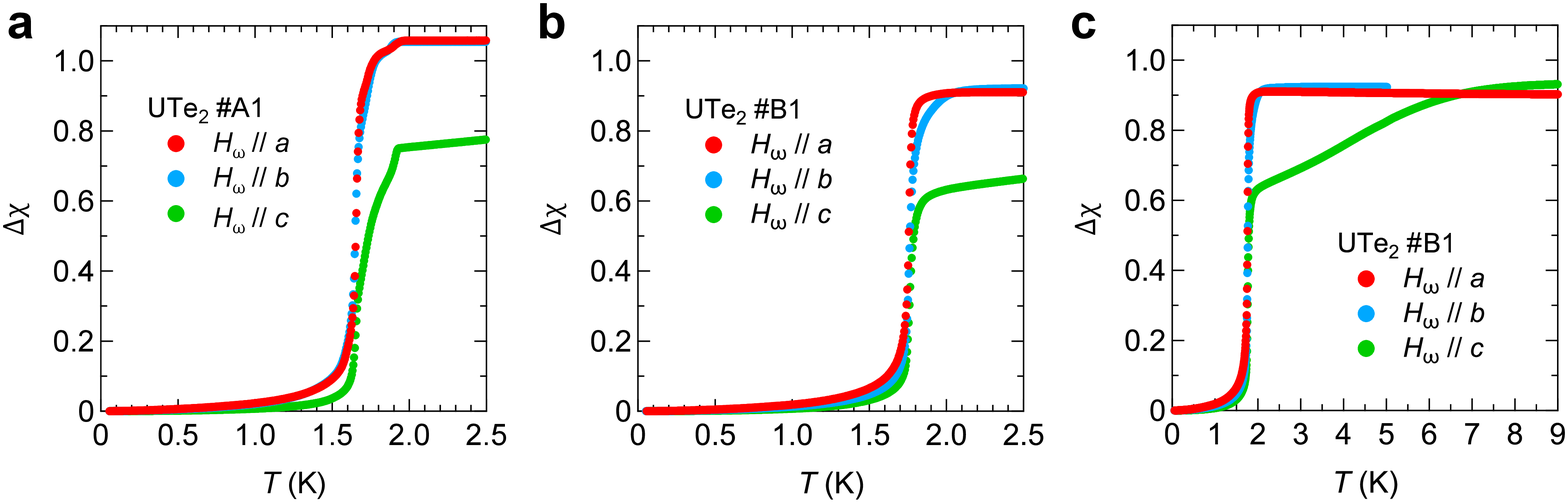}
\caption{{\bf Magnetic susceptibility along three crystallographic axes.} {\bf a,b,} Magnetic susceptibility shift $\Delta \chi \equiv \chi(T) - \chi(0)$ below 2.5\,K calculated from $\Delta f$ through Eq.\,\ref{f_chi} for crystals \#A1 ({\bf a}) and \#B1 ({\bf b}). {\bf c,} $\Delta \chi (T)$ up to 9\,K ($H_{\omega} \parallel a$ and $c$) and up to 5\,K ($H_{\omega} \parallel b$) for crystal \#B1.}
\label{chi}
\end{figure}

The frequency shift of the oscillator $\Delta f(T) \equiv f(T) - f(0)$ and ac magnetic susceptibility shift $\Delta \chi(T) \equiv \chi(T) - \chi(0)$ satisfy the relation\,\cite{Prozorov2000},
\begin{equation}
\frac{\Delta f(T)}{f_0} = -\frac{V_s}{2V_c(1-N)} \Delta \chi(T),
\label{f_chi}
\end{equation}
where $f_0=13.8$\,MHz is the resonant frequency without the sample, $V_s$ and $V_c$ are the sample and coil volumes, and $N$ is the demagnetization factor calculated from the equation\,\cite{Prozorov2018}
\begin{equation}
\frac{1}{1-N} = 1+\frac{4ab}{3c(a+b)}.
\end{equation}
$\Delta \chi_i$ data for crystals \#A1 and \#B1 with the magnetic field along $i$-axis calculated from the above equations are depicted in Figs.\,S\ref{chi}a,b, respectively. It is clearly seen that the values of $\Delta \chi_a (T_c)$ and $\Delta \chi_b (T_c)$ are nearly equal to 1, while $\Delta \chi_c (T_c)$ is about 0.8. This difference can be understood in consideration of the skin effect. The skin depth $\delta$ is calculated by
\begin{equation}
\delta = \sqrt{\frac{\rho}{\pi f_0 \mu_r \mu_0}},
\end{equation}
where $\rho$ is the resistivity, $f$ is the frequency of the ac magnetic field, and $\mu_0$ and $\mu_r$ are vacuum and relative permeability, respectively. By substituting $\rho=30$\,$\mu \Omega$cm (Fig.\,S\ref{Specific_Resis}c) and $\mu_r=1$, we obtain $\delta=75$\,$\mu$m. When the magnetic field is parallel to the $a$- or $b$-axes, the skin depth is larger than the sample thickness, and the skin effect can be neglected. Then, the total shift of $\Delta \chi_a (T_c) \approx \Delta \chi_b (T_c) \approx 1$ are expected in the perfect diamagnetic state. As shown in Fig.\,S6, $\Delta\chi_a(T_c)\approx \Delta\chi_b(T_c)\approx 1.05$ for crystal \#A1 and 0.9 for crystal \#B1, from which we can evaluate that the errors of the calculated $N$ values are within 10\%. On the other hand, when the magnetic field is parallel to the $c$-axis, the skin effect expels the magnetic field even in the normal state, which reduces the $\Delta\chi_c(T_c)$ value from unity as shown in Fig.\,S6. For crystal \#A1, $\Delta \chi_c(T_c) \approx 0.8$ is expected for $\delta=75 \,\mu$m, which is consistent with the experimental value obtained from crystal \#R1 picked up from the same batch as crystals \#A1 and \#A2. For crystal \#B1, we measured $\Delta \chi_i (T)$ up to 9\,K (Fig.\,S\ref{chi}c). Since the skin depth becomes longer and compatible to the sample thickness with increasing temperature, $\Delta\chi_c$ reaches almost the same value of $\Delta\chi_a$ and $\Delta\chi_b$ above 7\,K as shown in Fig.\,S6c. These results indicate that our estimation of the demagnetization factors $N$ is appropriate and $\Delta \chi (T)$ data are successfully derived from $\Delta f(T)$ data in crystals \#A1 and \#B1. We note that we measured $\Delta f(T)$ in another crystal (\#A2) (see Fig.\,3b and Fig.\,S\ref{sample2}), but the complicated sample shape of crystal \#A2 makes the quantitative analysis of $\Delta \lambda_i$ difficult. We have carried out the same analysis for crystal \#A2 as for crystal \#A1. However, the values of $\Delta\chi_{a,b}(T_c)$ for crystal \#A2 are much larger than unity, indicating that the demagnetization factor $N$ for crystal \#A2 is not properly obtained probably because of the complicated sample shape. Hence, we did not carry out further analysis for crystal \#A2.

In general, for a superconductor with anisotropic penetration depth components, $\chi_i$ can be expressed by using $\lambda_j$ and $\lambda_k$ through the equations \,\cite{Prozorov_arxiv},
\begin{numcases}
{}
1+\chi_a = \frac{\lambda_b}{R_a^c} + \frac{\lambda_c}{R_a^b} & \\
1+\chi_b = \frac{\lambda_c}{R_b^a} + \frac{\lambda_a}{R_b^c} & \\
1+\chi_c = \frac{\lambda_a}{R_c^b} + \frac{\lambda_b}{R_c^a},&
\end{numcases}
where $R_i^j$ and $R_i^k$ are the effective lengths when the magnetic field is applied along the $i$-axis.
For a 3D thin disk of radius $w$ and thickness $2d$ where $\lambda$ is isotropic in the in-plane directions, the previous study\,\cite{Prozorov2000} derived the effective length
\begin{equation}
R_{3D}=\frac{w}{2\left\{ 1+\left[ 1+\left( \frac{2d}{w} \right) ^2 \right] \arctan \left( \frac{w}{2d} \right) - \frac{2d}{w} \right\}}
\end{equation}
which satisfies
\begin{equation}
-\chi=1-\frac{\lambda}{R_{3D}} \tanh \left( \frac{R_{3D}}{\lambda} \right).
\end{equation}
For the development of the above discussions to an anisotropic case, we derive the anisotropic effective lengths 
\begin{eqnarray}
R_k^i = \frac{4i}{3\left\{ 1+\left[ 1+\left( \frac{2d}{w} \right) ^2 \right] \arctan \left( \frac{w}{2d} \right) - \frac{2d}{w} \right\}} \\
R_k^j = \frac{4j}{3\left\{ 1+\left[ 1+\left( \frac{2d}{w} \right) ^2 \right] \arctan \left( \frac{w}{2d} \right) - \frac{2d}{w} \right\}},
\end{eqnarray}
where $d=k$ and $w=8ij/3(i+j)$ for sample dimensions of $2a\times2b\times2c$ (here, $(i,j,k)=(a,b,c)$, $(b,c,a)$, or $(c,a,b)$), by assuming $1/R_{3D}=1/R_k^i+1/R_k^j$ and $R_k^i:R_k^j=i:j$. From the above discussions, we can derive $\Delta \lambda_a$, $\Delta \lambda_b$, and $\Delta \lambda_c$ through the relations,
\begin{equation}
R\Delta \lambda_i = -\frac{R_k^j R_i^k}{R_k^i} \Delta \chi_i + \frac{R_j^i R_k^j R_i^k}{R_k^i R_i^j} \Delta \chi_j + R_k^j \Delta \chi_k,
\end{equation}
where $R=1+R_b^a R_c^b R_a^c / R_b^c R_a^b R_c^a$. The calculated $\Delta \lambda_a$, $\Delta \lambda_b$, and $\Delta \lambda_c$ data for crystal \#B1 are shown in Fig.\,4b in the main text.

Since crystal \#A1 is a plate-like sample, we consider the demagnetization effect only for the field parallel to the $c$-axis direction and we use the approximations $R^b_a = b$, $R^c_a = c$, $R^a_b = a$, and $R^c_b = c$, for the following reasons. Because Eq.\,(S18) is derived by using an approximation\,\cite{Prozorov2000}
\begin{equation}
\frac{1}{1-N} = 1+\frac{w}{2d}
\end{equation}
which is usually used for a thin cylindrical sample, the above discussions are valid only if the sample shape can be treated as an effective cylindrical shape. Considering that the penetrating volume is proportional to the perimeter, the radius of the effective cylinder can be $\tilde{w}=2(i+j)/\pi$ when $H_{\omega} \parallel k$\,\cite{Prozorov2021}. Then, comparing Eq.\,(S13) and Eq.\,(S23) under the condition that $w=\tilde{w}$, we can find that the difference of these approximations becomes larger than 14\,\% for crystal \#A1 with $H_{\omega} \parallel a$ or $b$, while the difference is less than 3.2\,\% for crystal \#B1. On the other hand, Ref.\,\cite{Prozorov2021} shows that the effective dimension with $N<0.15$ in a cylindrical crystal has at most 10\,\% of difference from the crystal dimensions. Thus, we consider that, for crystal \#A1 with $H_{\omega} \parallel a$ or $H_{\omega} \parallel b$ where $N<0.15$ from Eq.\,(S13), it is better to use the crystal dimensions ($R^b_a = b$, $R^c_a = c$, $R^a_b = a$, and $R^c_b = c$) than the effective dimensions described in Eqs.\,(S20, S21). The obtained $\Delta \lambda_i (T)$ data for crystal \#A1 are shown in the main text. We stress that the robustness of our analyses against the errors of the effective dimensions has been confirmed by changing the values by $\pm 30$\,\% in section VIII.
\\

\section{Fitting-range dependence of exponent values and possibility of deep gap minima}

\begin{figure}[h]
\includegraphics[width=\linewidth]{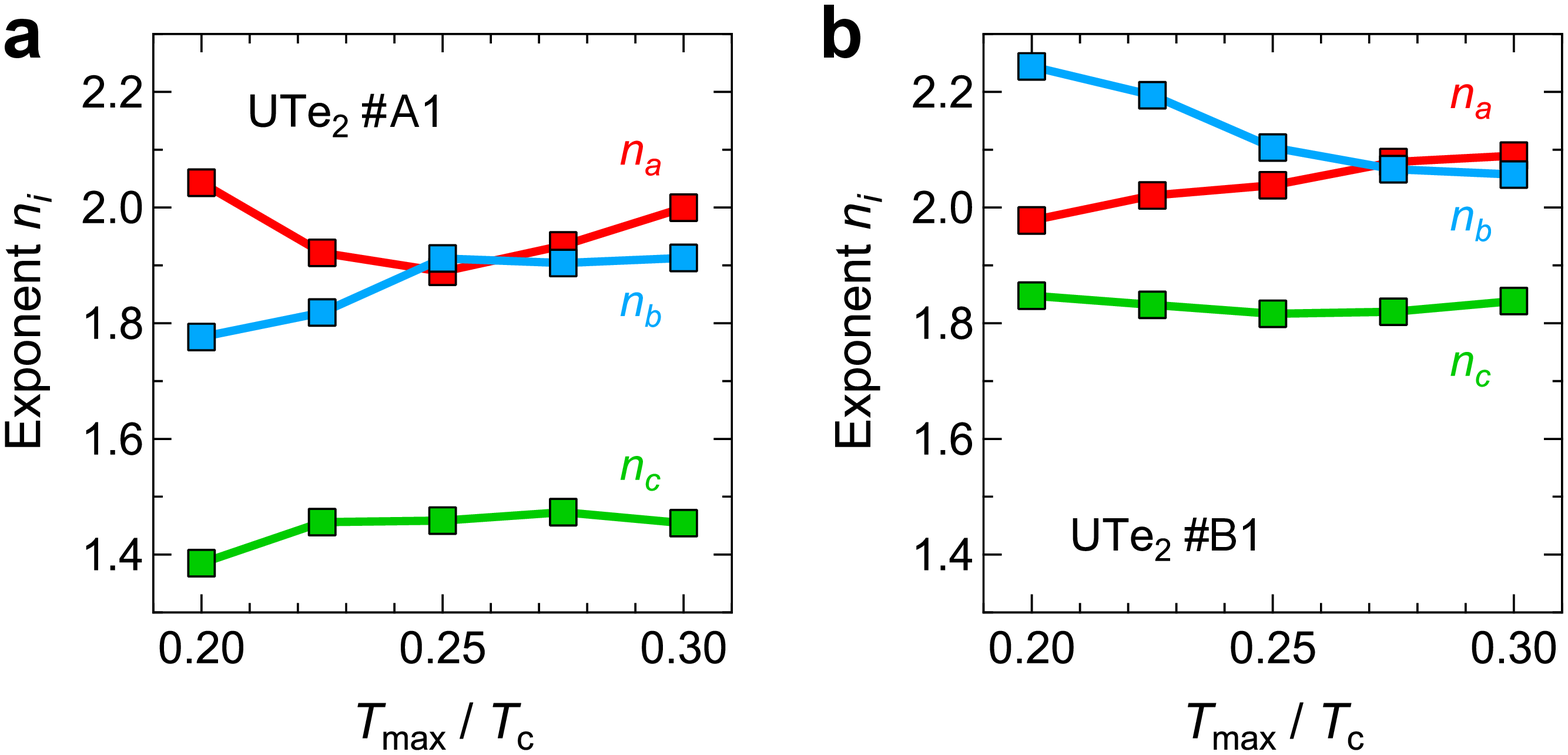}
\caption{{\bf Fitting-range dependence of the exponent values.} {\bf a,b,} Exponent values $n_i$ for crystals \#A1 ({\bf a}) and  \#B1 ({\bf b}) as a function of the maximum value of the fitting range $T_{\rm max}$.}
\label{Exponent}
\end{figure}

We checked the fitting range $T_{\rm max}$ dependence of the exponent values $n_i$ in crystals \#A1 and \#B1 obtained by the power-law fitting, $\Delta \lambda_i \propto T^{n_i}$ (see Fig.\,S\ref{Exponent}). We confirmed that our discussions on the gap structure via $n_i$ values are not affected by the $T_{\rm max}$ values. 
We have excluded the possibility of a line node gap in UTe$_2$ in section III.  However, deep gap minima can be possible, for example, if $c_2$ and $c_6$ are much larger than the other coefficients in Eq.\,\ref{B3u_iAu}, but such a possibility can be ruled out by the following experimental facts. When the size of the gap minima is comparable or larger than the experimental lowest temperature, the exponent obtained from the power-law fitting becomes higher as the maximum value $T_{\rm{max}}$ of the fitting range gets lower \cite{Cho2016,Ishihara2021}. In contrast, our experimental data do not show such a fitting-range dependence as shown in Fig.\,S\ref{Exponent}, indicating that the gap minima, if exist, should be really deep, much lower than the lowest temperature of our measurements. Even in the case of such deep minima, one can get important information from the sample dependence. In general, when the impurity scattering becomes large, the averaging effect of gap anisotropy leads to the increase of minimum gap, which results in a larger exponent value of low-temperature penetration depth. From the skin depth analysis in the normal state just above $T_c$, we find that the residual resistivity is $\sim2$ times higher in crystal \#A1 than \#B1, indicating larger impurity scattering in crystal \#A1 (see section VI). However, we observed lower exponents in the power-law analysis of penetration depth especially for the $c$-axis data, which is the opposite to the expected behavior of deep gap minima case. On the other hand, the observed tendency can be explained by the slightly different parameter in our proposed chiral order parameter case as discussed in section IX.
\\

\section{Robustness of analyses}

\begin{figure}[h]
\includegraphics[width=\linewidth]{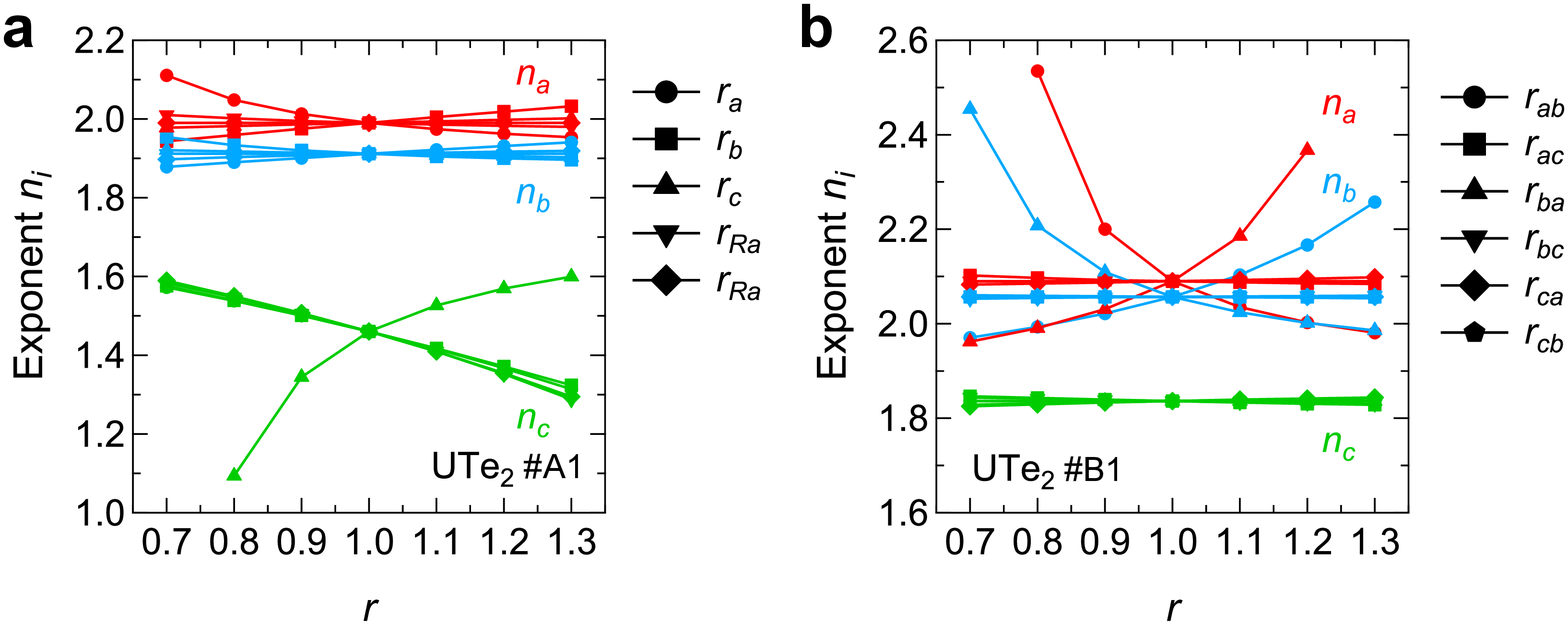}
\caption{{\bf Effective length dependence of the exponent values.} {\bf a,b,} Exponent values obtained from the power-law fitting to $\Delta \lambda_i (T)$ as a function of $r$ in crystals \#A1 ({\bf a}) and \#B1 ({\bf b}).}
\label{r_dependence}
\end{figure}

It is known that the effective dimensions $R_{3D}$ (Eq.\,S18) have usually errors of $\pm 20$\,\% even in the isotropic case\,\cite{Prozorov2000}. Thus, we need to check the effect of the errors on our analyses, especially on the exponent values obtained by the power-law fitting. To this end, we varied the effective dimensions as $a\rightarrow r_a a$, $b\rightarrow r_b b$, $c\rightarrow r_c c$, $R_c^b \rightarrow r_{bc} R_c^b$, and $R_c^a \rightarrow r_{ac} R_c^a$ for crystal \#A1 and $R_b^a \rightarrow r_{ab} R_b^a$, $R_c^a \rightarrow r_{ac} R_c^a$, $R_a^b \rightarrow r_{ba} R_a^b$, $R_c^b \rightarrow r_{bc} R_c^b$, $R_a^c \rightarrow r_{ca} R_a^c$, and $R_b^c \rightarrow r_{cb} R_b^c$ for crystal \#B1. The $r$ dependence of the exponent values $n_i$ are summarized in Fig.\,S\ref{r_dependence}. It can be seen that $n_a$ and $n_b$ in crystal \#A1 and $n_c$ in crystal \#B1 are almost independent on $r$. We emphasize here that $n_c$ is always lower than $n_a$ and $n_b$ in both crystals, which strongly confirms the presence of nearby point nodes near the $k_z$-axis as discussed in the next section.
\\

\section{Interference effect of point nodes}
\begin{figure}[tbp]
\includegraphics[width=0.8\linewidth]{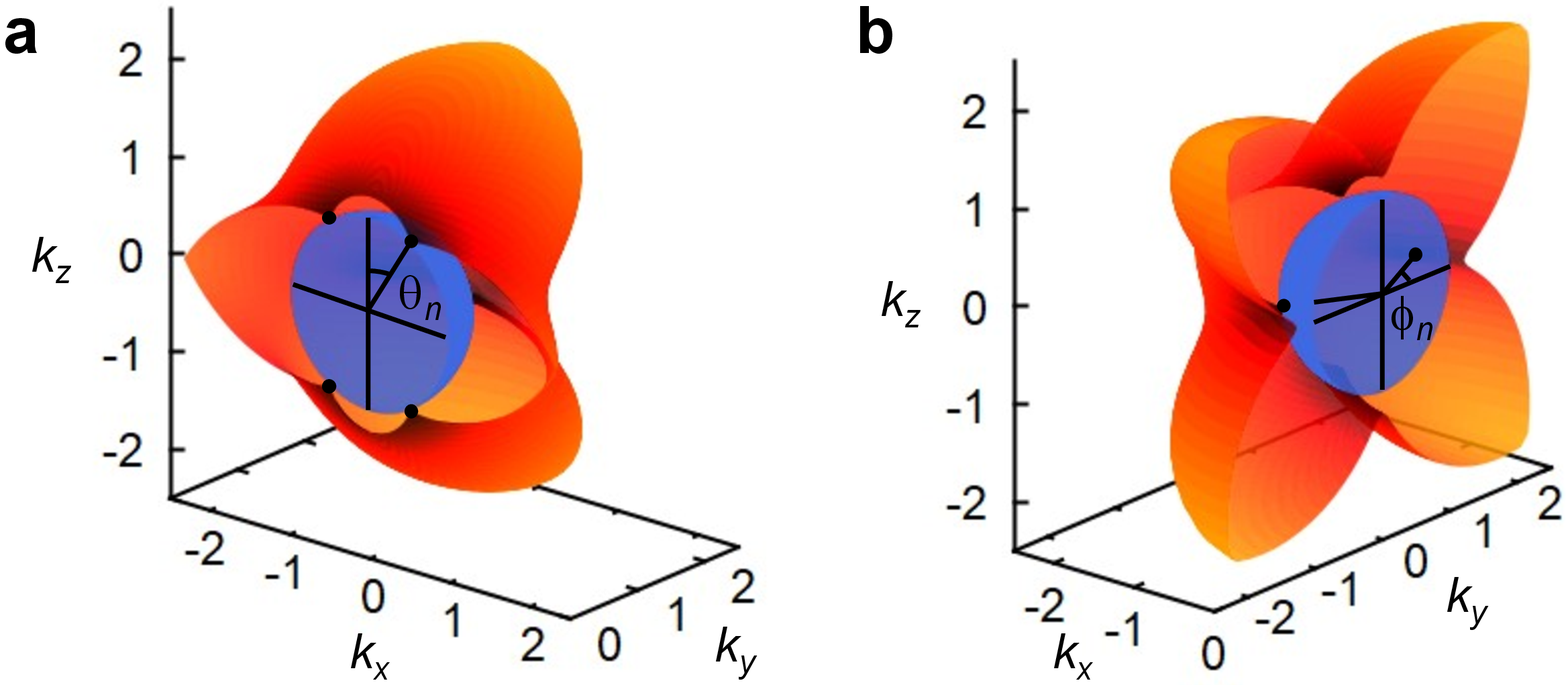}
\caption{{\bf Gap structure of the $B_{3u}+iA_u$ state.} {\bf a,b,} Angular dependence of the $B_{3u}+iA_u$ gap function and definitions of $\theta_n$ ({\bf a}) and $\phi_n$ ({\bf b}), respectively.}
\label{gap}
\end{figure}

In this section, we calculate $\Delta \lambda_i (T)/\lambda_i (0)$ with a simplified model for the $B_{3u}+iA_u$ state. We consider the order parameter
\begin{equation}
{\bm d} = \left(
\begin{array}{c}
k_x k_y k_z + ik_x \\
c_2 k_z + ik_y \\
c_3 k_y + ik_z
\end{array}
\right),
\end{equation}
which is obtained from Eq.\,\ref{B3u_iAu} for $c_1=c_4=c_5=c_6=1$. The positions of the point nodes are described by
\begin{equation}
\left\{
\begin{aligned}
& k_y=0 \\
& k_x=\pm \sqrt{c_2^2-1} k_z
\end{aligned}
\right. \qquad {\rm and} \qquad \left\{
\begin{aligned}
& k_z=0 \\
& k_x=\pm \sqrt{c_3^2-1} k_y
\end{aligned}
\right.,
\end{equation}
when $c_2>1$ and $c_3>1$. Here, we define the positions of point nodes by the angles $\theta_n$ and $\phi_n$ where $\tan (\theta_n) = \sqrt{c_2^2-1}$ and $\tan (\phi_n) = \sqrt{c_3^2-1}$ (Figs.\,S\ref{gap}a,b). The superfluid density along the $a$-, $b$-, and $c$-axes are calculated from
\begin{equation}
\left\{
\begin{aligned}
& \rho_a^b = 1-\frac{3}{4\pi T} \int_0^1 (1-z^2) \int_0^{2\pi} \left(
\begin{aligned}
& \cos^2 (\phi) \\
& \sin^2 (\phi)
\end{aligned}
\right) \int_0^{\infty} \cosh^{-2} \left( \frac{\sqrt{\epsilon^2+\Delta_0^2 (T) \hat{g}^2 (\theta, \phi)}}{2T} \right) {\rm d}\epsilon {\rm d}\theta {\rm d}\phi \\
& \rho_c = 1-\frac{3}{4\pi T} \int_0^1 z^2 \int_0^{2\pi} \int_0^{\infty} \cosh^{-2} \left( \frac{\sqrt{\epsilon^2+\Delta_0^2 (T) \hat{g}^2 (\theta, \phi)}}{2T} \right) {\rm d}\epsilon {\rm d}\theta {\rm d}\phi
\end{aligned}
\right.,
\end{equation}
where $z=\cos (\theta)$, $\Delta_0 (T)$ is the $T$ dependence of the gap size, and $\hat{g} (\theta, \phi)$ is the angular dependence of the gap function with a maximum value of unity\,\cite{Prozorov_review}. 
In this study, we used an approximation\,\cite{Kogan2021}
\begin{equation}
\Psi (t) = \frac{\pi T_c e^{-\left< \Omega^2 \rm{ln} \left| \Omega \right| \right>}}{e^\gamma} \rm{tanh} \left( e^\gamma \sqrt{\frac{8(1-t)}{7\zeta(3)t}} \frac{e^{\left< \Omega^2 \rm{ln} \left| \Omega \right| \right>}}{\sqrt{\left< \Omega^4 \right>}} \right)
\end{equation}
and $\Delta_0 (T) \hat{g}(\theta, \phi) = \Psi(T) \Omega(\theta, \phi)$, where $\left< \Omega^2 \right> = 1$, $\gamma \approx 0.577$ is the Euler constant, and $\left< \cdots \right>$ represents averaging over the Fermi surface.
From $\rho_i (T)$, we can calculate the anisotropic magnetic penetration depth $\Delta \lambda_i (T) / \lambda (0) = 1/\sqrt{\rho_i (T)} -1$.

\begin{figure}[tbp]
\includegraphics[width=0.8\linewidth]{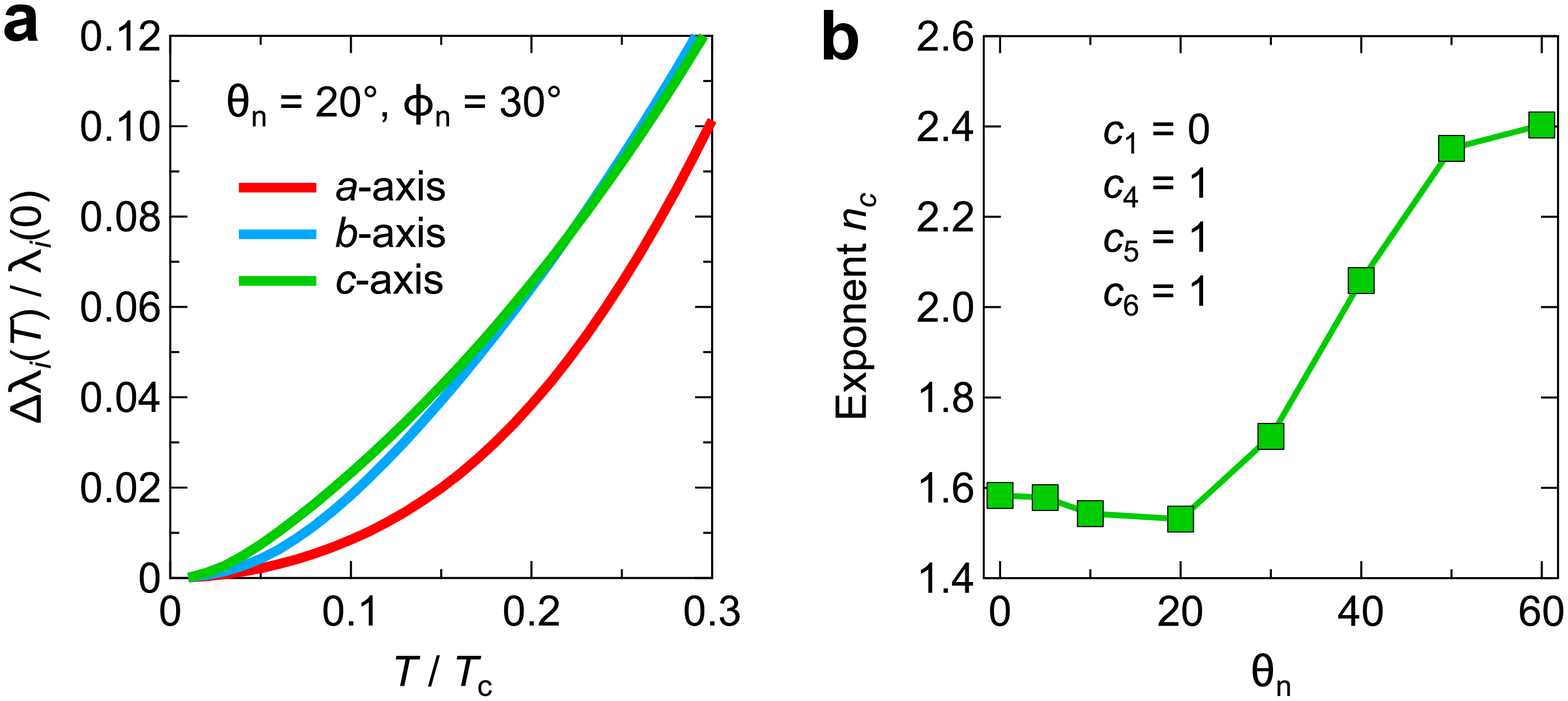}
\caption{{\bf Calculated anisotropic penetration depth and exponent values.} {\bf a,} Calculated $\Delta \lambda_i (T)/ \lambda_i (0)$ along the $i=a$, $b$, and $c$ axes with point nodes in the $\theta_n = 20^\circ$ and $\phi_n = 30^\circ$ directions. {\bf b,} Exponent values obtained from the power-law fitting $\Delta \lambda_c (T)/ \lambda_c (0) \propto T^{n_c}$ as a function of $\theta_n$.}
\label{Delta_lambda}
\end{figure}

First, we examine the nodal position dependence of $\Delta \lambda_i (T)$ in the $B_{3u}+iA_u$ state by changing the parameters $c_2$ and $c_3$ for the fixed parameters $c_1=0$, $c_4=1$, $c_5=1$, and $c_6=1$. As an example, the calculated $\Delta \lambda_i (T) / \lambda_i (0)$ for $\theta_n =20^\circ$ and $\phi_n =30^\circ$ are depicted in Fig.\,S\ref{Delta_lambda}a. We obtained the exponent value $n_c$ as a function of $\theta_n$, which reflects the closeness of the nearby point nodes, by power-law fitting up to $0.3T_c$ (Fig.\,S\ref{Delta_lambda}b). In this calculation, we find that the exponent value $n_c$ approaches 1.6 as $\theta_n \rightarrow 0$, and $n_c < 2$ is robust when the node angle $\theta_n$ is less than $20^\circ$. These results can be interpreted as an interference effect due to nearby two point nodes, in which each point node can no longer be treated as an individual. Thus, the observation of $n_c < 2$ for crystals \#A1 and \#B1 is indicative of the closeness of two point nodes.

\begin{figure}[tbp]
\includegraphics[width=0.9\linewidth]{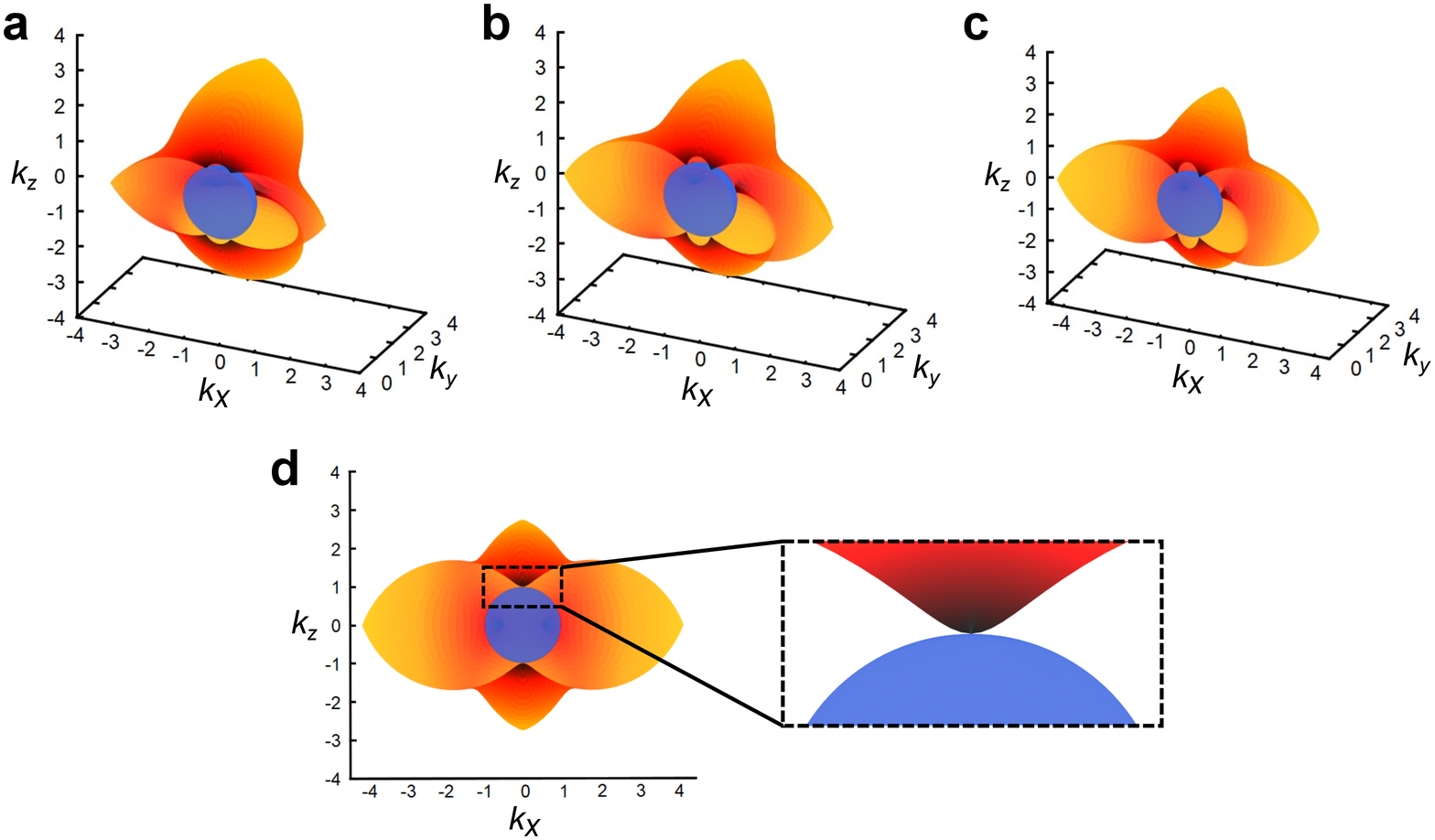}
\caption{{\bf $c_4$ value dependence of the gap structure in the $B_{3u}+iA_u$ state.} {\bf a-c,} Angular dependence of the $B_{3u}+iA_u$ gap function with $c_4=1.0$ ({\bf a}), $c_4=1.5$ ({\bf b}), and $c_4=2.0$ ({\bf c,d}). The other parameters are fixed as $c_1=0$, $c_5=1$, $c_6=1$, $c_2=\sqrt{c_6^2+c_4^2/{\rm tanh}^2 (\pi-\theta_n)}$, and $c_3=\sqrt{c_5^2+c_4^2/{\rm tanh}^2 (\pi-\phi_n)}$ for $\theta_n=15^\circ$ and $\phi_n=30^\circ$. {\bf d,} The same calculation with $c_4=2.0$ and $\theta_n=0$.}
\label{c4_dep_gap}
\end{figure}

\begin{figure}[tbp]
\includegraphics[width=\linewidth]{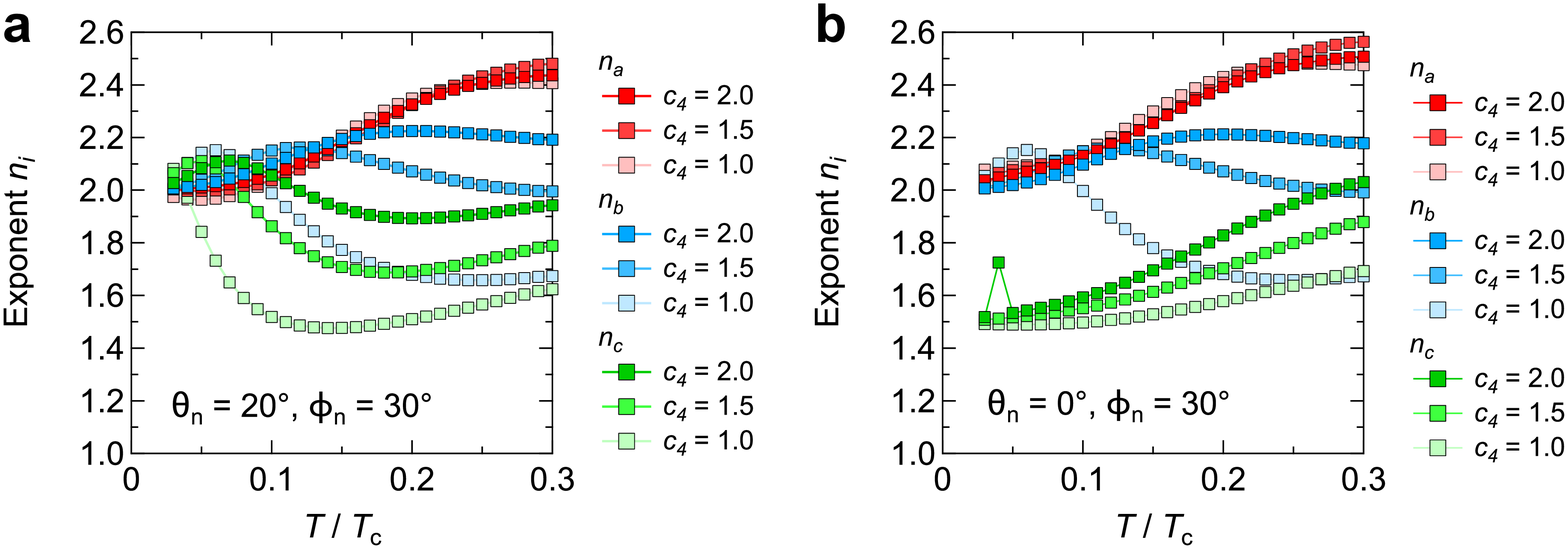}
\caption{{\bf Temperature dependent exponent values in the $B_{3u}+iA_u$ state.} {\bf a,b,} Temperature dependent exponents obtained by Eq.\,\ref{n_T_cal} by varying the $c_4$ value while fixing the angles $\theta_n = 20^\circ$ and $\phi_n = 30^\circ$ ({\bf a}) and $\theta_n = 0^\circ$ and $\phi_n = 30^\circ$ ({\bf b}) .}
\label{c4_dep_exponent}
\end{figure}

\begin{figure}[tbp]
\includegraphics[width=\linewidth]{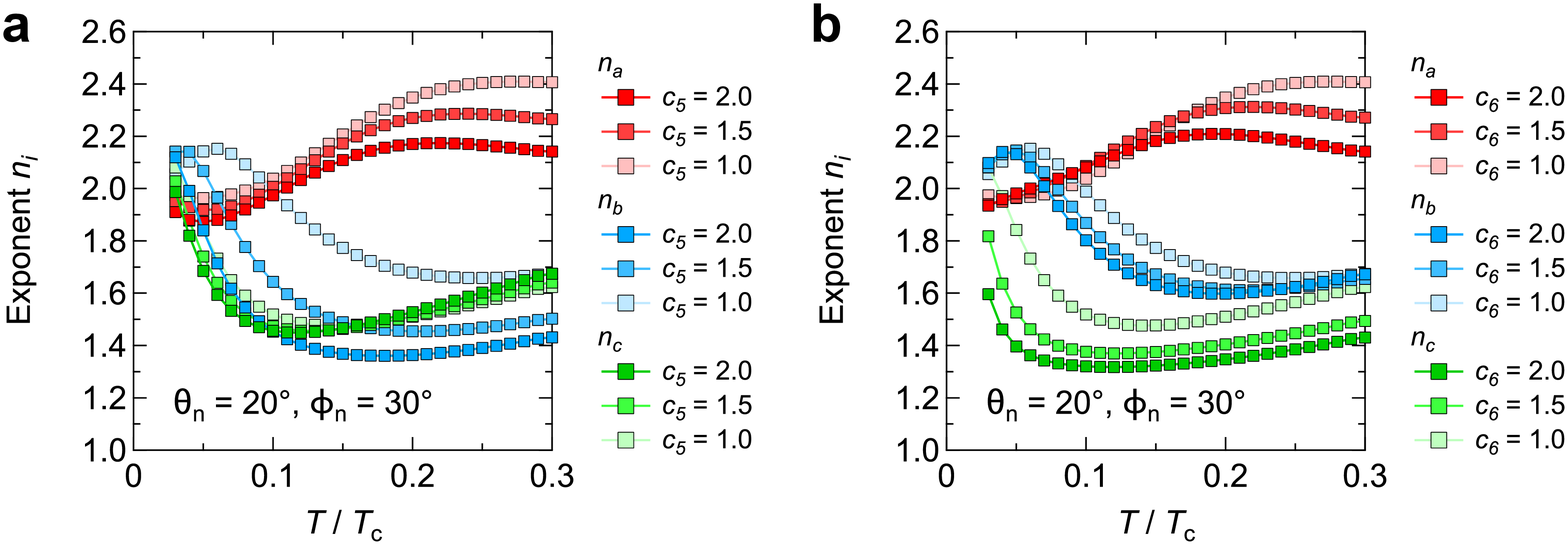}
\caption{{\bf $c_5$ and $c_6$ dependence of the exponent values.} {\bf a,b,} Temperature dependent exponents obtained by Eq.\,\ref{n_T_cal} by varying the $c_5$ and $c_6$ value while fixing the angles $\theta_n = 20^\circ$ and $\phi_n = 30^\circ$.}
\label{c5_c6_dep}
\end{figure}

Next, we investigate the temperature dependent exponent values by changing the parameters of $B_{3u}+iA_u$ state in Eq.\,\ref{B3u_iAu}. The gap structures change as shown in Fig.\,S\ref{c4_dep_gap} when the $c_4$ value is varied (here, we fixed $c_1=0$, $c_5=1$, $c_6=1$, $c_2=\sqrt{c_6^2+c_4^2/{\rm tanh}^2 (\pi-\theta_n)}$, and $c_3=\sqrt{c_5^2+c_4^2/{\rm tanh}^2 (\pi-\phi_n)}$ for $\theta_n = 15^\circ$ and $\phi_n = 30^\circ$). The temperature dependence of the exponents $n_i (T/T_c)$ can be calculated from
\begin{equation}
n_i = \frac{d \left[ {\rm ln} \left( \Delta \lambda_i (T/T_c) \right) \right]}{d \left[ {\rm ln} \left( T/T_c \right) \right]}
\label{n_T_cal}
\end{equation}
and the results are summarized in Figs.\,S\ref{c4_dep_exponent} and S\ref{c5_c6_dep}. We find that for all the parameters we use, $n_c$ is smaller than 2 in the temperature range above $\sim0.1T_c$, where we can analyze our experimental data, whereas it starts approaching 2 below $\sim0.1T_c$ except for $n_c\sim1.5$ at $\theta_n =0^\circ$ (Fig.\,S\ref{c4_dep_exponent}b) where the dispersion of the gap function near the point nodes is not linear but quadratic (Fig.\,S\ref{c4_dep_gap}d). This low-temperature recovery of the power of 2 is simply because the nearby point nodes can be treated as individual nodes in the $T\rightarrow 0$ limit. The crossover energy from $T^2$ to lower exponent region is determined by the maximum gap size between the nearby point nodes, and thus it depends on the locations of nodes and gap sizes of the $B_{3u}$ and $A_u$ components.

In order to clarify how robust the $B_{3u}+iA_u$ order parameter is in UTe$_2$, we compare the results of crystals \#A1 and \#B1 with  slightly different $T_c$s. We find that crystal \#B1 shows relatively larger exponent values along all the crystal directions than crystal \#A1 (see Fig.\,4), but the key feature that $n_c < 2$ and $n_a \sim 2$ remains for both crystals, which can be consistently explained by the $B_{3u}+iA_u$ state, considering the $c_4$ dependence of the exponent values as follows. 
Since we have already found that the exponent values are not sensitive to the nodal positions when $\theta_n$ is sufficiently small (Fig.\,S\ref{c4_dep_exponent}), we should focus on the changes of $c_4$, $c_5$, or $c_6$ values. Then, keeping the nodal positions and varying the parameters $c_4$, $c_5$, and $c_6$ independently, we find that, while larger $c_5$ and $c_6$ values provide smaller $n_b$ and $n_c$ in the whole temperature range (Fig.\,S\ref{c5_c6_dep}), larger $c_4$ values enlarge the $n_b$ and $n_c$ values (Fig.\,S\ref{c4_dep_exponent}), which is consistent with the experimental results. This $c_4$ dependence can be interpreted as a weakening of the interference effect caused by a larger gap size between the two point nodes (Fig.\,S\ref{c4_dep_gap}). Thus, we can conclude that the $d_x$ component of the $A_u$ order parameter may be an origin of the sample-dependent exponent values. However, for detailed understanding of the sample dependence, further studies are desired.


We comment on the calculated $n_a$ and $n_b$ values which are slightly different from the observed $n_a \approx n_b \approx 2$. The reason for this difference may be related to the FS geometry. As mentioned in the main text, the recent ARPES study\,\cite{Miao2020_SI} shows a 3D FS around the $Z$ point and quasi-2D FSs. Then, we can expect that $\Delta \lambda_a (T)/ \lambda_a (0)$ and $\Delta \lambda_b (T)/ \lambda_b (0)$ are mainly contributed from the point nodes on the quasi-2D FSs, while $\Delta \lambda_c (T)/ \lambda_c (0)$ is mainly contributed from the point nodes on the 3D FS. Therefore, our calculations based on a spherical FS lead to the calculated $n_a$ and $n_b$ values different from the experimental values. Furthermore, we note that the angles $\theta_n$ and $\phi_n$ are determined by the coefficients of the order parameter. Thus, for more precise calculations of $n_a$ and $n_b$, experimental studies on the FS geometry, such as quantum oscillations, and the sizes of the order parameter components are awaited.
\\

\section{Estimation of $\lambda(0)$}
In general, the anisotropies of $\lambda(0)$ and $\xi(T_c)$ can be given by 
$\gamma_{\lambda}^2(0)=\lambda_j^2(0)/\lambda_i^2(0)=\langle v_{F,i}^2 \rangle/\langle v_{F,j}^2 \rangle$ and $\gamma_{\xi}^2(T_c)=\xi_i^2(T_c)/\xi_j^2(T_c)=\langle \Omega^2 v_{F,i}^2 \rangle/\langle \Omega^2 v_{F,j}^2 \rangle$, respectively, where $\Omega$ describes the angular dependence of the gap function \cite{Kogan2019}. Thus, the anisotropy of $\lambda(0)$ cannot be estimated from the anisotropy of $\xi(T_c)$ when the superconducting gap function has a strong momentum dependence. However, the correction is not that large: In the recent calculations for the gap structure with two point nodes (like the $B_{3u}$ state) on a spherical Fermi surface, the ratio of $\gamma_\xi(T_c)$ to $\gamma_\lambda(0)$ is as small as $\sqrt{2}${\,}\cite{Kogan2021}, which does not change the main results that the $a$-axis quasiparticle excitations are significantly small compared with those along other directions (see new Fig.\,4).
 
Although previous theoretical studies have calculated the anisotropies of $\lambda$ and $\xi$ for superconductors with anisotropic gap functions\,\cite{Kogan2019}, theoretical calculations for a heavy fermion superconductor is challenging because of the highly renormalized band structure. Furthermore, an experimental estimation of $\lambda_i(0)$ from $H_{\rm{c1}}$ measurements is also challenging in UTe$_2$ because of the field dependent pairing strength \cite{Paulsen2021}. In this study, therefore, we simply ignored the anisotropy of gap function for the relation between the anisotropies of coherence length and penetration depth. As a result, we obtained $\lambda_a (0)\sim \lambda_c(0) > \lambda_b(0)$, which is consistent with the band structure calculations that the bands along the $a$- and $c$-axes consist of the $d$- and $f$-orbitals of U, while the band along the $b$-axis consists of the $p$-orbitals of Te (see e.g., Refs.\,[11-13,20,S26]). Note that this does not affect the exponent analysis of the low-temperature penetration depth.

If we consider the effect of gap anisotropy $\Omega$ in the proposed $B_{3u}+iA_u$ case, we expect that the ratio $\left< \Omega^2 v_{F,i}^2 \right> / \left< v_{F,i}^2 \right>$ is largest for $a$-axis, near which nodes are absent. This immediately implies that in comparison with the isotropic approximation we use, the actual ratio $\xi_i^2 (T_c)/\lambda_i^2 (0)$ should be larger for $a$-axis, and thus the $\lambda_a (0)$ value is expected to be even larger. Then, the $a$-axis data $\Delta \lambda_a(T)/\lambda_a(0)$ should be even smaller, and thus the anisotropy of low-energy quasiparticle excitations becomes more pronounced. From these reasons, we argue that the results of this simple analysis are meaningful, and the conclusions obtained are valid.
\\

\clearpage

\begin{table}[tbp]
\centering
\caption{\bf Basis functions and nodal types for even-parity order parameters in the point group $D_{2h}$.}
\label{ST1}
\begin{tabular*}{0.4\columnwidth}{@{\extracolsep{\fill}}ccc}
\hline\hline
IR & Basis functions & Nodes  \\ \hline
$A_{1g}$ & $k_x^2, k_y^2, k_z^2$ & None  \\
$B_{1g}$ & $k_x k_y$ & Line  \\
$B_{2g}$ & $k_z k_x$ & Line  \\
$B_{3g}$ & $k_x k_y$ & Line  \\
\hline\hline
\end{tabular*}
\end{table}

\clearpage

\begin{thebibliography}{10}
\expandafter\ifx\csname url\endcsname\relax
  \def\url#1{\texttt{#1}}\fi
\expandafter\ifx\csname urlprefix\endcsname\relax\def\urlprefix{URL }\fi
\providecommand{\bibinfo}[2]{#2}
\providecommand{\eprint}[2][]{\url{#2}}

\bibitem{Kallin2016}
\bibinfo{author}{Kallin, C.} \& \bibinfo{author}{Berlinsky, J.}
\newblock \bibinfo{title}{Chiral superconductors}.
\newblock \emph{\bibinfo{journal}{Rep. Prog. Phys.}}
  \textbf{\bibinfo{volume}{79}}, \bibinfo{pages}{054502}
  (\bibinfo{year}{2016}).
\newblock \urlprefix\url{https://doi.org/10.1088/0034-4885/79/5/054502}.

\bibitem{Kozii2016}
\bibinfo{author}{Kozii, V.}, \bibinfo{author}{Venderbos, J. W.~F.} \&
  \bibinfo{author}{Fu, L.}
\newblock \bibinfo{title}{{Three-dimensional Majorana fermions in chiral
  superconductors}}.
\newblock \emph{\bibinfo{journal}{Sci. Adv.}} \textbf{\bibinfo{volume}{2}}
  (\bibinfo{year}{2016}).
\newblock
  \urlprefix\url{https://advances.sciencemag.org/content/2/12/e1601835}.

\bibitem{Ran2019_S}
\bibinfo{author}{Ran, S.} \emph{et~al.}
\newblock \bibinfo{title}{Nearly ferromagnetic spin-triplet superconductivity}.
\newblock \emph{\bibinfo{journal}{Science}} \textbf{\bibinfo{volume}{365}},
  \bibinfo{pages}{684--687} (\bibinfo{year}{2019}).
\newblock \urlprefix\url{https://science.sciencemag.org/content/365/6454/684}.

\bibitem{Aoki2019}
\bibinfo{author}{Aoki, D.} \emph{et~al.}
\newblock \bibinfo{title}{Unconventional superconductivity in heavy fermion
  {UTe$_2$}}.
\newblock \emph{\bibinfo{journal}{J. Phys. Soc. Jpn.}}
  \textbf{\bibinfo{volume}{88}}, \bibinfo{pages}{043702}
  (\bibinfo{year}{2019}).
\newblock \urlprefix\url{https://doi.org/10.7566/JPSJ.88.043702}.

\bibitem{Ran2019_NP}
\bibinfo{author}{Ran, S.} \emph{et~al.}
\newblock \bibinfo{title}{Extreme magnetic field-boosted superconductivity}.
\newblock \emph{\bibinfo{journal}{Nat. Phys.}} \textbf{\bibinfo{volume}{15}},
  \bibinfo{pages}{1250--1254} (\bibinfo{year}{2019}).
\newblock \urlprefix\url{https://doi.org/10.1038/s41567-019-0670-x}.

\bibitem{Nakamine2021}
\bibinfo{author}{Nakamine, G.} \emph{et~al.}
\newblock \bibinfo{title}{Anisotropic response of spin susceptibility in the
  superconducting state of {UTe$_2$} probed with {$^{125}$Te-NMR} measurement}.
\newblock \emph{\bibinfo{journal}{Phys. Rev. B}}
  \textbf{\bibinfo{volume}{103}}, \bibinfo{pages}{L100503}
  (\bibinfo{year}{2021}).
\newblock
  \urlprefix\url{https://link.aps.org/doi/10.1103/PhysRevB.103.L100503}.

\bibitem{Jiao2020}
\bibinfo{author}{Jiao, L.} \emph{et~al.}
\newblock \bibinfo{title}{Chiral superconductivity in heavy-fermion metal
  {UTe$_2$}}.
\newblock \emph{\bibinfo{journal}{Nature}} \textbf{\bibinfo{volume}{579}},
  \bibinfo{pages}{523--527} (\bibinfo{year}{2020}).
\newblock \urlprefix\url{https://doi.org/10.1038/s41586-020-2122-2}.

\bibitem{Hayes_arxiv}
\bibinfo{author}{Hayes, I.~M.} \emph{et~al.}
\newblock \bibinfo{title}{Multicomponent superconducting order parameter in {UTe$_2$}}.
\newblock \emph{\bibinfo{journal}{Science}} \textbf{\bibinfo{volume}{373}},
  \bibinfo{pages}{797--801} (\bibinfo{year}{2021}).
\newblock \urlprefix\url{https://www.science.org/doi/abs/10.1126/science.abb0272}.

\bibitem{Bae2021}
\bibinfo{author}{Bae, S.} \emph{et~al.}
\newblock \bibinfo{title}{Anomalous normal fluid response in a chiral
  superconductor {UTe$_2$}}.
\newblock \emph{\bibinfo{journal}{Nat. Commun.}} \textbf{\bibinfo{volume}{12}},
  \bibinfo{pages}{2644} (\bibinfo{year}{2021}).
\newblock \urlprefix\url{https://doi.org/10.1038/s41467-021-22906-6}.

\bibitem{Metz2019}
\bibinfo{author}{Metz, T.} \emph{et~al.}
\newblock \bibinfo{title}{Point-node gap structure of the spin-triplet
  superconductor {UTe$_2$}}.
\newblock \emph{\bibinfo{journal}{Phys. Rev. B}}
  \textbf{\bibinfo{volume}{100}}, \bibinfo{pages}{220504}
  (\bibinfo{year}{2019}).
\newblock \urlprefix\url{https://link.aps.org/doi/10.1103/PhysRevB.100.220504}.

\bibitem{Ishizuka2021}
\bibinfo{author}{Ishizuka, J.} \& \bibinfo{author}{Yanase, Y.}
\newblock \bibinfo{title}{Periodic anderson model for magnetism and
  superconductivity in {UTe$_2$}}.
\newblock \emph{\bibinfo{journal}{Phys. Rev. B}}
  \textbf{\bibinfo{volume}{103}}, \bibinfo{pages}{094504}
  (\bibinfo{year}{2021}).
\newblock \urlprefix\url{https://link.aps.org/doi/10.1103/PhysRevB.103.094504}.

\bibitem{Nevidomskyy_arxiv}
\bibinfo{author}{Nevidomskyy, A.~H.}
\newblock \bibinfo{title}{Stability of a nonunitary triplet pairing on the
  border of magnetism in {UTe$_2$}}.
\newblock \emph{\bibinfo{journal}{preprint}} \bibinfo{pages}{arXiv:2001.02699}
  (\bibinfo{year}{2020}).

\bibitem{Shishidou2021}
\bibinfo{author}{Shishidou, T.}, \bibinfo{author}{Suh, H.~G.},
  \bibinfo{author}{Brydon, P. M.~R.}, \bibinfo{author}{Weinert, M.} \&
  \bibinfo{author}{Agterberg, D.~F.}
\newblock \bibinfo{title}{Topological band and superconductivity in {UTe$_2$}}.
\newblock \emph{\bibinfo{journal}{Phys. Rev. B}}
  \textbf{\bibinfo{volume}{103}}, \bibinfo{pages}{104504}
  (\bibinfo{year}{2021}).
\newblock \urlprefix\url{https://link.aps.org/doi/10.1103/PhysRevB.103.104504}.

\bibitem{Knebel2019}
\bibinfo{author}{Knebel, G.} \emph{et~al.}
\newblock \bibinfo{title}{Field-reentrant superconductivity close to a
  metamagnetic transition in the heavy-fermion superconductor {UTe$_2$}}.
\newblock \emph{\bibinfo{journal}{J. Phys. Soc. Jpn.}}
  \textbf{\bibinfo{volume}{88}}, \bibinfo{pages}{063707}
  (\bibinfo{year}{2019}).
\newblock \urlprefix\url{https://doi.org/10.7566/JPSJ.88.063707}.

\bibitem{Kittaka2020}
\bibinfo{author}{Kittaka, S.} \emph{et~al.}
\newblock \bibinfo{title}{Orientation of point nodes and nonunitary triplet
  pairing tuned by the easy-axis magnetization in {UTe$_2$}}.
\newblock \emph{\bibinfo{journal}{Phys. Rev. Res.}}
  \textbf{\bibinfo{volume}{2}}, \bibinfo{pages}{032014} (\bibinfo{year}{2020}).
\newblock
  \urlprefix\url{https://link.aps.org/doi/10.1103/PhysRevResearch.2.032014}.

\bibitem{Thomas2020}
\bibinfo{author}{Thomas, S.~M.} \emph{et~al.}
\newblock \bibinfo{title}{Evidence for a pressure-induced antiferromagnetic
  quantum critical point in intermediate-valence {UTe$_2$}}.
\newblock \emph{\bibinfo{journal}{Sci. Adv.}} \textbf{\bibinfo{volume}{6}}
  (\bibinfo{year}{2020}).
\newblock
  \urlprefix\url{https://advances.sciencemag.org/content/6/42/eabc8709}.
\newblock
  \eprint{https://advances.sciencemag.org/content/6/42/eabc8709.full.pdf}.

\bibitem{Braithwaite2019}
\bibinfo{author}{Braithwaite, D.} \emph{et~al.}
\newblock \bibinfo{title}{Multiple superconducting phases in a nearly
  ferromagnetic system}.
\newblock \emph{\bibinfo{journal}{Commun. Phys.}} \textbf{\bibinfo{volume}{2}},
  \bibinfo{pages}{147} (\bibinfo{year}{2019}).
\newblock \urlprefix\url{https://doi.org/10.1038/s42005-019-0248-z}.

\bibitem{Aoki2020}
\bibinfo{author}{Aoki, D.} \emph{et~al.}
\newblock \bibinfo{title}{Multiple superconducting phases and unusual
  enhancement of the upper critical field in {UTe$_2$}}.
\newblock \emph{\bibinfo{journal}{J. Phys. Soc. Jpn.}}
  \textbf{\bibinfo{volume}{89}}, \bibinfo{pages}{053705}
  (\bibinfo{year}{2020}).
\newblock \urlprefix\url{https://doi.org/10.7566/JPSJ.89.053705}.

\bibitem{Haga_arxiv}
\bibinfo{author}{Haga, Y.} \emph{et~al.}
\newblock \bibinfo{title}{Effect of uranium deficiency on normal and superconducting properties in unconventional superconductor {UTe$_2$}}.
\newblock \emph{\bibinfo{journal}{preprint}} \bibinfo{pages}{arXiv:2112.13468}
  (\bibinfo{year}{2021}).

\bibitem{Thomas2021}
\bibinfo{author}{Thomas, S.~M.} \emph{et~al.}
\newblock \bibinfo{title}{Spatially inhomogeneous superconductivity in ${\mathrm{UTe}}_{2}$}.
\newblock \emph{\bibinfo{journal}{Phys. Rev. B}}
  \textbf{\bibinfo{volume}{104}}, \bibinfo{pages}{224501}
  (\bibinfo{year}{2021}).
\newblock
  \urlprefix\url{https://link.aps.org/doi/10.1103/PhysRevB.104.224501}.

\bibitem{Rosa_arxiv}
\bibinfo{author}{Rosa, P. F.~S.} \emph{et~al.}
\newblock \bibinfo{title}{Single-component superconducting state in {UTe$_2$}
  at 2\,{K}}.
\newblock \emph{\bibinfo{journal}{preprint}} \bibinfo{pages}{arXiv:2110.06200}
  (\bibinfo{year}{2021}).

\bibitem{Miao2020}
\bibinfo{author}{Miao, L.} \emph{et~al.}
\newblock \bibinfo{title}{Low energy band structure and symmetries of {UTe$_2$}
  from angle-resolved photoemission spectroscopy}.
\newblock \emph{\bibinfo{journal}{Phys. Rev. Lett.}}
  \textbf{\bibinfo{volume}{124}}, \bibinfo{pages}{076401}
  (\bibinfo{year}{2020}).
\newblock
  \urlprefix\url{https://link.aps.org/doi/10.1103/PhysRevLett.124.076401}.

\bibitem{Ishizuka2019}
\bibinfo{author}{Ishizuka, J.}, \bibinfo{author}{Sumita, S.},
  \bibinfo{author}{Daido, A.} \& \bibinfo{author}{Yanase, Y.}
\newblock \bibinfo{title}{Insulator-metal transition and topological
  superconductivity in {UTe$_2$} from a first-principles calculation}.
\newblock \emph{\bibinfo{journal}{Phys. Rev. Lett.}}
  \textbf{\bibinfo{volume}{123}}, \bibinfo{pages}{217001}
  (\bibinfo{year}{2019}).
\newblock
  \urlprefix\url{https://link.aps.org/doi/10.1103/PhysRevLett.123.217001}.

\bibitem{Paulsen_arxiv}
\bibinfo{author}{Paulsen, C.} \emph{et~al.}
\newblock \bibinfo{title}{Anomalous anisotropy of the lower critical field and
  Meissner effect in {UTe$_2$}}.
\newblock \emph{\bibinfo{journal}{Phys. Rev. B}}
  \textbf{\bibinfo{volume}{103}}, \bibinfo{pages}{L180501}
  (\bibinfo{year}{2021}).
\newblock
  \urlprefix\url{https://link.aps.org/doi/10.1103/PhysRevB.103.L180501}.

\bibitem{Aoki_review}
\bibinfo{author}{Aoki, D.}, \bibinfo{author}{Ishida, K.} \&
  \bibinfo{author}{Flouquet, J.}
\newblock \bibinfo{title}{Review of {U-based} ferromagnetic superconductors:
  {Comparison between UGe$_2$, URhGe, and UCoGe}}.
\newblock \emph{\bibinfo{journal}{J. Phys. Soc. Jpn.}}
  \textbf{\bibinfo{volume}{88}}, \bibinfo{pages}{022001}
  (\bibinfo{year}{2019}).
\newblock \urlprefix\url{https://doi.org/10.7566/JPSJ.88.022001}.

\bibitem{Gross1986}
\bibinfo{author}{Gro{\ss}, F.} \emph{et~al.}
\newblock \bibinfo{title}{Anomalous temperature dependence of the magnetic
  field penetration depth in superconducting {UBe$_{13}$}}.
\newblock \emph{\bibinfo{journal}{Z. Phys. B}} \textbf{\bibinfo{volume}{64}},
  \bibinfo{pages}{175--188} (\bibinfo{year}{1986}).

\bibitem{Sigrist1991}
\bibinfo{author}{Sigrist, M.}, \& \bibinfo{author}{Ueda, K.},
\newblock \bibinfo{title}{Phenomenological theory of unconventional superconductivity}.
\newblock \emph{\bibinfo{journal}{Rev. Mod. Phys.}}
  \textbf{\bibinfo{volume}{63}}, \bibinfo{pages}{239}
  (\bibinfo{year}{1991}).
\newblock
  \urlprefix\url{https://link.aps.org/doi/10.1103/RevModPhys.63.239}.

\bibitem{Hirschfeld1993}
\bibinfo{author}{Hirschfeld, P.~J.} \& \bibinfo{author}{Goldenfeld, N.}
\newblock \bibinfo{title}{Effect of strong scattering on the low-temperature
  penetration depth of a d-wave superconductor}.
\newblock \emph{\bibinfo{journal}{Phys. Rev. B}} \textbf{\bibinfo{volume}{48}},
  \bibinfo{pages}{4219--4222} (\bibinfo{year}{1993}).
\newblock \urlprefix\url{https://link.aps.org/doi/10.1103/PhysRevB.48.4219}.

\bibitem{Hashimoto2013}
\bibinfo{author}{Hashimoto, K.} \emph{et~al.}
\newblock \bibinfo{title}{Anomalous superfluid density in quantum critical
  superconductors}.
\newblock \emph{\bibinfo{journal}{Proc. Natl. Acad. Sci. USA}}
  \textbf{\bibinfo{volume}{110}}, \bibinfo{pages}{3293--3297}
  (\bibinfo{year}{2013}).
\newblock \urlprefix\url{https://www.pnas.org/content/110/9/3293}.

\bibitem{Kosztin1997}
\bibinfo{author}{Kosztin, I.} \& \bibinfo{author}{Leggett, A.~J.}
\newblock \bibinfo{title}{Nonlocal effects on the magnetic penetration depth in
  $d$-wave superconductors}.
\newblock \emph{\bibinfo{journal}{Phys. Rev. Lett.}}
  \textbf{\bibinfo{volume}{79}}, \bibinfo{pages}{135--138}
  (\bibinfo{year}{1997}).
\newblock \urlprefix\url{https://link.aps.org/doi/10.1103/PhysRevLett.79.135}.

\bibitem{Eo_arxiv}
\bibinfo{author}{Eo, Y.~S.} \emph{et~al.}
\newblock \bibinfo{title}{Anomalous $c$-axis transport response of
  {UTe$_{2}$}}.
\newblock \emph{\bibinfo{journal}{preprint}} \bibinfo{pages}{arXiv:2101.03102}
  (\bibinfo{year}{2021}).

\bibitem{Prozorov_review}
\bibinfo{author}{Prozorov, R.} \& \bibinfo{author}{Giannetta, R.~W.}
\newblock \bibinfo{title}{Magnetic penetration depth in unconventional
  superconductors}.
\newblock \emph{\bibinfo{journal}{Supercond. Sci. Technol.}}
  \textbf{\bibinfo{volume}{19}}, \bibinfo{pages}{R41--R67}
  (\bibinfo{year}{2006}).
\newblock \urlprefix\url{https://doi.org/10.1088/0953-2048/19/8/r01}.

\bibitem{Prozorov_arxiv}
\bibinfo{author}{Prozorov, R.}
\newblock \bibinfo{title}{{Meissner-London} state in anisotropic
  superconductors of cuboidal shape}.
\newblock \emph{\bibinfo{journal}{preprint}} \bibinfo{pages}{arXiv:2101.06489}
  (\bibinfo{year}{2021}).

\end{thebibliography}

\begin{thebibliography}{10}
\expandafter\ifx\csname url\endcsname\relax
  \def\url#1{\texttt{#1}}\fi
\expandafter\ifx\csname urlprefix\endcsname\relax\def\urlprefix{URL }\fi
\providecommand{\bibinfo}[2]{#2}
\providecommand{\eprint}[2][]{\url{#2}}

\bibitem{Hayes_science}
\bibinfo{author}{Hayes, I.~M.} \emph{et~al.}
\newblock \bibinfo{title}{Multicomponent superconducting order parameter in {UTe$_2$}}.
\newblock \emph{\bibinfo{journal}{Science}} \textbf{\bibinfo{volume}{373}},
  \bibinfo{pages}{797--801} (\bibinfo{year}{2021}).
\newblock \urlprefix\url{https://www.science.org/doi/abs/10.1126/science.abb0272}.

\bibitem{Hirschfeld1993}
\bibinfo{author}{Hirschfeld, P.~J.} \& \bibinfo{author}{Goldenfeld, N.}
\newblock \bibinfo{title}{Effect of strong scattering on the low-temperature
  penetration depth of a $d$-wave superconductor}.
\newblock \emph{\bibinfo{journal}{Phys. Rev. B}} \textbf{\bibinfo{volume}{48}},
  \bibinfo{pages}{4219--4222} (\bibinfo{year}{1993}).
\newblock \urlprefix\url{https://link.aps.org/doi/10.1103/PhysRevB.48.4219}.

\bibitem{Hashimoto2013}
\bibinfo{author}{Hashimoto, K.} \emph{et~al.}
\newblock \bibinfo{title}{Anomalous superfluid density in quantum critical
  superconductors}.
\newblock \emph{\bibinfo{journal}{Proc. Natl. Acad. Sci. USA}}
  \textbf{\bibinfo{volume}{110}}, \bibinfo{pages}{3293--3297}
  (\bibinfo{year}{2013}).
\newblock \urlprefix\url{https://www.pnas.org/content/110/9/3293}.

\bibitem{Mizukami2016}
\bibinfo{author}{Mizukami, Y.} \emph{et~al.}
\newblock \bibinfo{title}{Evolution of quasiparticle excitations with enhanced
  electron correlations in superconducting {$A$Fe$_{2}$As$_{2}$ ($A=$ K, Rb,
  and Cs)}}.
\newblock \emph{\bibinfo{journal}{Phys. Rev. B}} \textbf{\bibinfo{volume}{94}},
  \bibinfo{pages}{024508} (\bibinfo{year}{2016}).
\newblock \urlprefix\url{https://link.aps.org/doi/10.1103/PhysRevB.94.024508}.

\bibitem{Kosztin1997}
\bibinfo{author}{Kosztin, I.} \& \bibinfo{author}{Leggett, A.~J.}
\newblock \bibinfo{title}{Nonlocal effects on the magnetic penetration depth in
  $d$-wave superconductors}.
\newblock \emph{\bibinfo{journal}{Phys. Rev. Lett.}}
  \textbf{\bibinfo{volume}{79}}, \bibinfo{pages}{135--138}
  (\bibinfo{year}{1997}).
\newblock \urlprefix\url{https://link.aps.org/doi/10.1103/PhysRevLett.79.135}.

\bibitem{Ran2019_SI}
\bibinfo{author}{Ran, S.} \emph{et~al.}
\newblock \bibinfo{title}{Nearly ferromagnetic spin-triplet superconductivity}.
\newblock \emph{\bibinfo{journal}{Science}} \textbf{\bibinfo{volume}{365}},
  \bibinfo{pages}{684--687} (\bibinfo{year}{2019}).
\newblock \urlprefix\url{https://science.sciencemag.org/content/365/6454/684}.

\bibitem{Prozorov2014}
\bibinfo{author}{Prozorov, R.} \emph{et~al.}
\newblock \bibinfo{title}{Effect of electron irradiation on superconductivity
  in single crystals of
  $\mathrm{Ba}({\mathrm{Fe}}_{1\ensuremath{-}x}{\mathrm{Ru}}_{x}{)}_{2}{\mathrm{As}}_{2}$
  ($x=0.24$)}.
\newblock \emph{\bibinfo{journal}{Phys. Rev. X}} \textbf{\bibinfo{volume}{4}},
  \bibinfo{pages}{041032} (\bibinfo{year}{2014}).
\newblock \urlprefix\url{https://link.aps.org/doi/10.1103/PhysRevX.4.041032}.

\bibitem{Takenaka2017}
\bibinfo{author}{Takenaka, T.} \emph{et~al.}
\newblock \bibinfo{title}{Full-gap superconductivity robust against disorder in
  heavy-fermion ${\mathrm{CeCu}}_{2}{\mathrm{Si}}_{2}$}.
\newblock \emph{\bibinfo{journal}{Phys. Rev. Lett.}}
  \textbf{\bibinfo{volume}{119}}, \bibinfo{pages}{077001}
  (\bibinfo{year}{2017}).
\newblock
  \urlprefix\url{https://link.aps.org/doi/10.1103/PhysRevLett.119.077001}.

\bibitem{Kim2015}
\bibinfo{author}{Kim, H.} \emph{et~al.}
\newblock \bibinfo{title}{Nodal to nodeless superconducting energy-gap
  structure change concomitant with fermi-surface reconstruction in the
  heavy-fermion compound ${\mathrm{CeCoIn}}_{5}$}.
\newblock \emph{\bibinfo{journal}{Phys. Rev. Lett.}}
  \textbf{\bibinfo{volume}{114}}, \bibinfo{pages}{027003}
  (\bibinfo{year}{2015}).
\newblock
  \urlprefix\url{https://link.aps.org/doi/10.1103/PhysRevLett.114.027003}.

\bibitem{Bonn1994}
\bibinfo{author}{Bonn, D.~A.} \emph{et~al.}
\newblock \bibinfo{title}{Comparison of the influence of Ni and Zn impurities on the electromagnetic properties of ${\mathrm{YBa}}_{2}$${\mathrm{Cu}}_{3}$${\mathrm{O}}_{6.95}$}.
\newblock \emph{\bibinfo{journal}{Phys. Rev. B}}
  \textbf{\bibinfo{volume}{50}}, \bibinfo{pages}{4051}
  (\bibinfo{year}{1994}).
\newblock
  \urlprefix\url{https://link.aps.org/doi/10.1103/PhysRevB.50.4051}.

\bibitem{Ishizuka2021_SI}
\bibinfo{author}{Ishizuka, J.} \& \bibinfo{author}{Yanase, Y.}
\newblock \bibinfo{title}{Periodic anderson model for magnetism and
  superconductivity in {UTe$_2$}}.
\newblock \emph{\bibinfo{journal}{Phys. Rev. B}}
  \textbf{\bibinfo{volume}{103}}, \bibinfo{pages}{094504}
  (\bibinfo{year}{2021}).
\newblock \urlprefix\url{https://link.aps.org/doi/10.1103/PhysRevB.103.094504}.

\bibitem{Sigrist_Ueda}
\bibinfo{author}{Sigrist, M.} \& \bibinfo{author}{Ueda, K.}
\newblock \bibinfo{title}{Phenomenological theory of unconventional
  superconductivity}.
\newblock \emph{\bibinfo{journal}{Rev. Mod. Phys.}}
  \textbf{\bibinfo{volume}{63}}, \bibinfo{pages}{239--311}
  (\bibinfo{year}{1991}).
\newblock \urlprefix\url{https://link.aps.org/doi/10.1103/RevModPhys.63.239}.

\bibitem{Aoki_review_SI}
\bibinfo{author}{Aoki, D.}, \bibinfo{author}{Ishida, K.} \&
  \bibinfo{author}{Flouquet, J.}
\newblock \bibinfo{title}{Review of {U-based} ferromagnetic superconductors:
  {Comparison between UGe$_2$, URhGe, and UCoGe}}.
\newblock \emph{\bibinfo{journal}{J. Phys. Soc. Jpn.}}
  \textbf{\bibinfo{volume}{88}}, \bibinfo{pages}{022001}
  (\bibinfo{year}{2019}).
\newblock \urlprefix\url{https://doi.org/10.7566/JPSJ.88.022001}.

\bibitem{Nevidomskyy_arxiv_SI}
\bibinfo{author}{Nevidomskyy, A.~H.}
\newblock \bibinfo{title}{Stability of a nonunitary triplet pairing on the
  border of magnetism in {UTe$_2$}}.
\newblock \emph{\bibinfo{journal}{preprint}} \bibinfo{pages}{arXiv:2001.02699}
  (\bibinfo{year}{2020}).

\bibitem{Prozorov2000}
\bibinfo{author}{Prozorov, R.}, \bibinfo{author}{Giannetta, R.~W.},
  \bibinfo{author}{Carrington, A.} \& \bibinfo{author}{Araujo-Moreira, F.~M.}
\newblock \bibinfo{title}{Meissner-london state in superconductors of
  rectangular cross section in a perpendicular magnetic field}.
\newblock \emph{\bibinfo{journal}{Phys. Rev. B}} \textbf{\bibinfo{volume}{62}},
  \bibinfo{pages}{115--118} (\bibinfo{year}{2000}).
\newblock \urlprefix\url{https://link.aps.org/doi/10.1103/PhysRevB.62.115}.

\bibitem{Prozorov2018}
\bibinfo{author}{Prozorov, R.} \& \bibinfo{author}{Kogan, V.~G.}
\newblock \bibinfo{title}{Effective demagnetizing factors of diamagnetic
  samples of various shapes}.
\newblock \emph{\bibinfo{journal}{Phys. Rev. Applied}}
  \textbf{\bibinfo{volume}{10}}, \bibinfo{pages}{014030}
  (\bibinfo{year}{2018}).
\newblock
  \urlprefix\url{https://link.aps.org/doi/10.1103/PhysRevApplied.10.014030}.

\bibitem{Prozorov_arxiv}
\bibinfo{author}{Prozorov, R.}
\newblock \bibinfo{title}{{Meissner-London} state in anisotropic
  superconductors of cuboidal shape}.
\newblock \emph{\bibinfo{journal}{preprint}} \bibinfo{pages}{arXiv:2101.06489}
  (\bibinfo{year}{2021}).
 
\bibitem{Prozorov2021}
\bibinfo{author}{Prozorov, R.}
\newblock \bibinfo{title}{Meissner-london susceptibility of superconducting
  right circular cylinders in an axial magnetic field}.
\newblock \emph{\bibinfo{journal}{Phys. Rev. Applied}}
  \textbf{\bibinfo{volume}{16}}, \bibinfo{pages}{024014}
  (\bibinfo{year}{2021}).
\newblock
  \urlprefix\url{https://link.aps.org/doi/10.1103/PhysRevApplied.16.024014}.

\bibitem{Cho2016}
\bibinfo{author}{Cho, K.} \emph{et~al.}
\newblock \bibinfo{title}{Energy gap evolution across the superconductivity dome in single crystals of (Ba$_{1-x}$K$_x$)Fe$_2$As$_2$}.\newblock \emph{\bibinfo{journal}{Sci. Adv.}}
  \textbf{\bibinfo{volume}{2}}, \bibinfo{pages}{e1600807}
  (\bibinfo{year}{2016}).
\newblock
  \urlprefix\url{https://link.aps.org/doi/10.1126/sciadv.1600807}.

\bibitem{Ishihara2021}
\bibinfo{author}{Ishihara, K.} \emph{et~al.}
\newblock \bibinfo{title}{Tuning the parity mixing of singlet-septet pairing in a half-Heusler superconductor}.
\newblock \emph{\bibinfo{journal}{Phys. Rev. X}}
  \textbf{\bibinfo{volume}{11}}, \bibinfo{pages}{041048}
  (\bibinfo{year}{2021}).
\newblock
  \urlprefix\url{https://link.aps.org/doi/10.1103/PhysRevApplied.16.024014}.
  
\bibitem{Prozorov_review}
\bibinfo{author}{Prozorov, R.} \& \bibinfo{author}{Giannetta, R.~W.}
\newblock \bibinfo{title}{Magnetic penetration depth in unconventional
  superconductors}.
\newblock \emph{\bibinfo{journal}{Supercond. Sci. Technol.}}
  \textbf{\bibinfo{volume}{19}}, \bibinfo{pages}{R41--R67}
  (\bibinfo{year}{2006}).
\newblock \urlprefix\url{https://doi.org/10.1088/0953-2048/19/8/r01}.

\bibitem{Kogan2021}
\bibinfo{author}{Kogan, V.~G.} \& \bibinfo{author}{Prozorov, R.}
\newblock \bibinfo{title}{Temperature dependence of London penetration depth
  anisotropy in superconductors with anisotropic order parameters}.
\newblock \emph{\bibinfo{journal}{Phys. Rev. B}}
  \textbf{\bibinfo{volume}{103}}, \bibinfo{pages}{054502}
  (\bibinfo{year}{2021}).
\newblock \urlprefix\url{https://link.aps.org/doi/10.1103/PhysRevB.103.054502}.

\bibitem{Miao2020_SI}
\bibinfo{author}{Miao, L.} \emph{et~al.}
\newblock \bibinfo{title}{Low energy band structure and symmetries of {UTe$_2$}
  from angle-resolved photoemission spectroscopy}.
\newblock \emph{\bibinfo{journal}{Phys. Rev. Lett.}}
  \textbf{\bibinfo{volume}{124}}, \bibinfo{pages}{076401}
  (\bibinfo{year}{2020}).
\newblock
  \urlprefix\url{https://link.aps.org/doi/10.1103/PhysRevLett.124.076401}.

\bibitem{Kogan2019}
\bibinfo{author}{Kogan, V.~G.} , \bibinfo{author}{Prozorov, R.} \& \bibinfo{author}{Koshelev, A.~E.}
\newblock \bibinfo{title}{Temperature-dependent anisotropies of upper critical field and London penetration depth}.
\newblock \emph{\bibinfo{journal}{Phys. Rev. B}}
  \textbf{\bibinfo{volume}{100}}, \bibinfo{pages}{014518}
  (\bibinfo{year}{2019}).
\newblock
  \urlprefix\url{https://link.aps.org/doi/10.1103/PhysRevB.100.014518}.

\bibitem{Paulsen2021}
\bibinfo{author}{Paulsen, C.} \emph{et~al.}
\newblock \bibinfo{title}{Anomalous anisotropy of the lower critical field and Meissner effect in UTe$_2$}.
\newblock \emph{\bibinfo{journal}{Phys. Rev. B}}
  \textbf{\bibinfo{volume}{103}}, \bibinfo{pages}{L180501}
  (\bibinfo{year}{2021}).
\newblock
  \urlprefix\url{https://link.aps.org/doi/10.1103/PhysRevB.103.L180501}.

\bibitem{Xu2019}
\bibinfo{author}{Xu, Y.} , \bibinfo{author}{Sheng, Y.} \& \bibinfo{author}{Yang, Y.-F.}
\newblock \bibinfo{title}{Quasi-Two-Dimensional Fermi Surfaces and Unitary Spin-Triplet Pairing in the Heavy Fermion Superconductor UTe$_2$}.
\newblock \emph{\bibinfo{journal}{Phys. Rev. Lett.}}
  \textbf{\bibinfo{volume}{123}}, \bibinfo{pages}{217002}
  (\bibinfo{year}{2019}).
\newblock
  \urlprefix\url{https://link.aps.org/doi/10.1103/PhysRevLett.123.217002}.

\end{thebibliography}
\end{document}